\documentclass{article}

\usepackage{arxiv}

\usepackage[utf8]{inputenc} 
\usepackage[T1]{fontenc}    
\usepackage{hyperref}       
\usepackage{url}            
\usepackage{booktabs}       
\usepackage{amsfonts}       
\usepackage{nicefrac}       
\usepackage{microtype}      
\usepackage{lipsum}
\usepackage{soul}
\usepackage{wrapfig}
\usepackage{diagbox}
\usepackage{caption} 
\usepackage[english]{babel}

\usepackage[linesnumbered,ruled,vlined,onelanguage]{algorithm2e}

\usepackage{amsmath}
\usepackage{graphicx,graphbox,psfrag,epsf}
\usepackage{subcaption}
\usepackage{float}
\usepackage[colorinlistoftodos]{todonotes}
\usepackage{enumerate}
\usepackage{natbib}
\usepackage{amssymb}
\usepackage{bbm}

\newcommand{\sub}{\mathbb{S}}
\newcommand{\euc}{\mathbb{X}}
\newcommand{\param}{\theta}
\newcommand{\tn}{\mathcal{N}_{\sub}}

\title{Flexible Mixture Modeling on Constrained Spaces}

\author{
  Putu Ayu~Sudyanti \\
  Department of Statistics\\
  Purdue University\\
  West Lafayette, IN 47906 \\
  \texttt{psudyant@purdue.edu} \\
   \And
   Vinayak~Rao\thanks{Corresponding author.} \\
  Department of Statistics\\
  Purdue University\\
  West Lafayette, IN 47906 \\
  \texttt{varao@purdue.edu} \\
}

\begin{document}
\maketitle

\begin{abstract}
This paper addresses challenges in flexibly modeling multimodal data that lie on constrained spaces. 
Such data are commonly found in spatial applications, such as climatology and criminology, where measurements are restricted to a geographical area. Other settings include domains where unsuitable recordings are discarded, such as flow-cytometry measurements. 
A simple approach to modeling such data is through the use of mixture models, especially nonparametric mixture models. 
Mixture models, while flexible and theoretically well-understood, are unsuitable for settings involving complicated constraints, leading to difficulties in specifying the component distributions and in evaluating normalization constants. Bayesian inference over the parameters of these models results in posterior distributions that are doubly-intractable. We address this problem via an algorithm based on rejection sampling and data augmentation. We view samples from a truncated distribution as outcomes of a rejection sampling scheme, where proposals are made from a simple mixture model and are rejected if they violate the constraints. Our scheme proceeds by imputing the rejected samples given mixture parameters and then resampling parameters given all samples. We study two modeling approaches: mixtures of truncated {Gaussians} and truncated mixtures of {Gaussians}, {along with their associated} Markov chain Monte Carlo sampling algorithms. We also discuss variations of the models, as well as approximations {to} improve mixing, reduce computational cost, and lower  variance. 
We present results on simulated data and apply our algorithms to two problems; one involving flow-cytometry data, and the other, crime recorded in the city of Chicago. 
\end{abstract}


\section{Introduction}
\label{sec:intro}
This paper studies approaches to flexibly modeling multimodal data that lie on constrained spaces. 
An example is crime data, where measurements are restricted to within the complex boundaries of a geographical entity, either because none exist outside (due to topographical features like water bodies) or because measurements {outside} belong to another city/state/country.
{Another example is flow cytometry}, where measurements with component-values outside some range (e.g.\ 0 to 1024) {are} discarded.  Other instances include single cell RNA-sequencing data~\citep{cao2017probabilistic} (where count-measurements below some threshold are discarded), operational risk modeling \citep{luo2009addressing} (where loss data only above a threshold are provided), stock price data \citep{aban2006parameter}, mortality~\citep{alai2013lifetime}, survival~\citep{Cain2011}, capture-recapture data~\citep{manning2010estimating}, where certain outcomes are truncated ~\citep{CensTrunc}, animal movement ~\citep{patterson2008state}, and climate data ~\citep{easterling2000climate}.  
In all of the above cases, it is important to accurately account for the boundaries of the constraint set, to avoid biases from boundary effects incorrectly interacting with smoothness assumptions inherent in typical probability models. Doing so, however, raises computational challenges due to the need to evaluate integrals over complicated subsets of a Euclidean space. This problem is exacerbated when the data exhibits rich multimodal and correlation structure, a common situation that requires mixture (and sometimes nonparametric mixture) models. 

In situations with simple constraints (such as the unit square), edge effects can be avoided by changing the parametric family used for the mixture components. For example, \cite{kottas2007bayesian} and \cite{matechou2017modelling} model the intensity function of an inhomogeneous Poisson process in a rectangular region $D$ using nonparametric mixtures of beta or gamma distributions. 
While such models allow multimodality and can accurately account for edge effects, they cannot easily model correlation structure in multi-dimensional data, something that is natural to Gaussian mixture models. Another approach is to transform the data to be unconstrained and then model the transformed data with a mixture of Gaussians. This can suffer from edge-effects that are not easy to characterize.
Importantly, both approaches are also not applicable to more complex constraint sets like the city of Chicago (Figure~\ref{fig:examples}, right).

We address this problem in this paper, developing and extending a methodology proposed in \citet{rao2016data} that avoids having to compute intractable terms arising from complex constraints.  
Our approach is to treat observations lying on the constrained space as the outcome of a rejection-sampling algorithm~\citep{Robert2005} with a proposal distribution $q$ defined on the simpler unconstrained space and with observations falling outside the constraint set discarded.  
Following \citet{rao2016data, beskos06}, we carry out inference over the distribution $q$ by imputing the rejected samples. Implicit in $q$ is all information about the original distribution of interest. 
Working directly with the unconstrained $q$ and imputing the rejected proposals allows us to use standard Bayesian modeling and computational techniques (such as nonparametric Bayesian models like Dirichlet process mixture models~\citep{lo1984class, EscWes1995}, and associated Gibbs samplers based on {the Chinese restaurant process} \citep{neal2000markov} and the stick-breaking process \citep{ishwaran2001gibbs}. 
In an application example in~\citep{rao2016data}, the authors briefly considered a setting that we call {\em truncated mixtures of Gaussians}. We study this in more detail, and also consider another natural approach: {\em mixtures of truncated {Gaussians}}. For both models, we describe exact Markov chain Monte Carlo (MCMC) schemes to impute the rejected proposals, allowing inference over cluster assignments, cluster parameters, and cluster weights.  In our experiments, we observe that naively implementing this can result in poor MCMC mixing. To speed up computations, improve mixing, and reduce MCMC estimation variance, we also propose and study modifications of the original models and the associated MCMC sampling algorithms. 
The resulting algorithms significantly outperform the original algorithms especially in higher-dimensional settings. 

We start by briefly reviewing the rejection sampling method and the associated data augmentation algorithm of~\citet{beskos06, rao2016data} in Section~\ref{sec:rejsamp}. 
In Section~\ref{sec:tmix} we define the two modeling approaches along with details of the associated MCMC samplers.
Section~\ref{sec:thrdars} outlines variations to {improve} mixing and computational efficiency.
Finally, in Section~\ref{sec:simstudy}, we present the results of our simulation studies on synthetic data, as well as real datasets of crime and flow cytometry measurements.

\section{Data augmentation for rejection sampling}
\label{sec:rejsamp}
\begin{figure}
\centering
\begin{minipage}{0.3\textwidth}
	\includegraphics[width=0.98\textwidth]{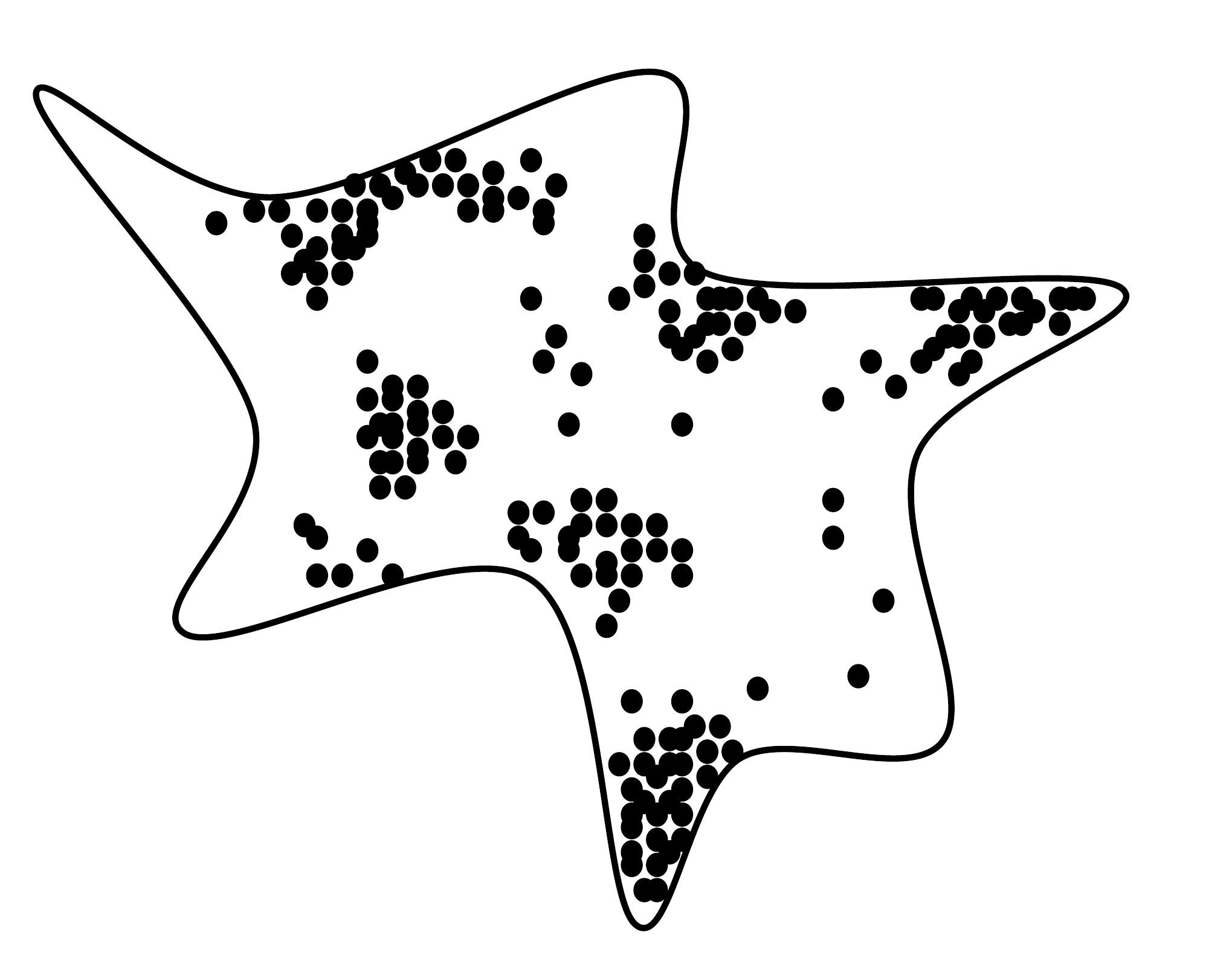}
    \end{minipage}
\begin{minipage}{0.3\textwidth}
	\includegraphics[width=0.75\textwidth]{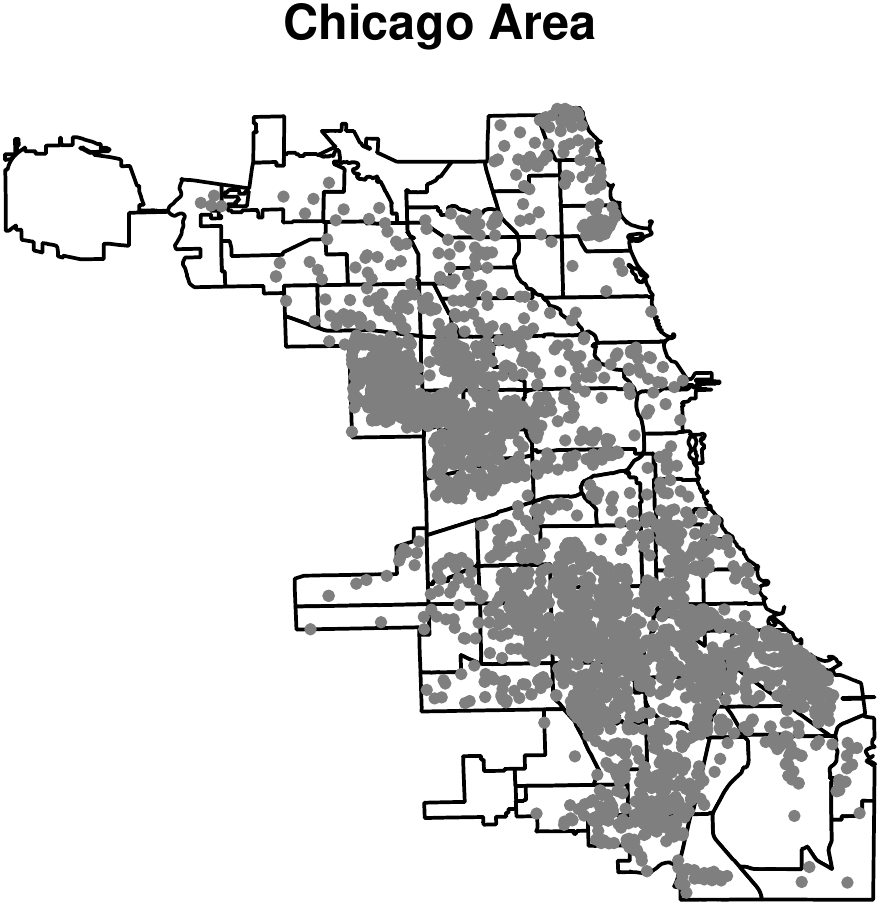}
    \end{minipage}
\begin{minipage}{0.38\textwidth}
\caption{Data lying on constrained subsets of Euclidean space. The left pane shows synthetic observations lying in the space $\sub$ specified by the 
solid black line. The right pane shows the location of homicide events in the city of Chicago. Here, the city limit defines the space $\sub$.} 
\label{fig:examples}
\end{minipage}
\vspace{-.1in}
\end{figure}

We consider observations $X=\{x_1,\ldots,x_n\}$ lying on a subset $\sub$ of a Euclidean space $\euc$. 
The set $\sub$ might be the set of positive reals, the unit sphere, or something more interesting like the city of Chicago. 
We show a simulated and a real example in Figure~\ref{fig:examples}. 
Our methodology is most useful for nonstandard sets like those in 
the figure. 
However, it is useful even for more common subsets like the unit square or the simplex. 
For these situations, standard approaches can be inadequate, either discarding correlation structure (e.g.\ mixtures of products of betas on the unit square) or imposing strong edge-effects (such as models involving transformed multivariate Gaussians). 
Our methodology provides a simple modeling framework for a wide variety of constraint sets where practitioners are only required to specify an indicator function for that set.

We model the observations as i.i.d.\ draws from a probability distribution $p(x)$ whose support equals to $\sub$. Our goal is to estimate $p(x)$ from observations $X$ and to do so, we will take a Bayesian approach by placing a prior over $p(x)$ and studying the resulting posterior distribution. 
Unfortunately, the requirement that $p(x)$ be restricted to $\sub$ raises challenges for both model specification as well as computation. For many interesting settings, the {probability density} $p(x)$ will involve an integral over $\sub$ and will therefore be intractable for all but simple choices of $\sub$. Posterior inference is consequently a doubly-intractable problem~\citep{murray2006}.

Our approach is to regard the distribution $p(x)$ as a restriction and renormalization on $\sub$ of some other distribution $q(x)$ on the ambient space $\euc$. Thus, $p(x) \propto \mathbbm{1}_\sub(x)q(x)$, where $\mathbbm{1}_\sub(\cdot)$ is the indicator function for $\sub$ and $q(x)$ is a standard distribution chosen such that MCMC posterior sampling techniques already exist. 
Defining $p(x)$ this way allows  sharp drops in the probability density from inside to outside $\sub$ and avoids undesirable smoothing effects at the boundaries. We restrict ourselves to subsets having positive probability under $q(x)$ and in practice, to subsets of $\euc$ having positive Lebesgue measure.
Our methodology does not apply to situations where $\sub$ is a lower-dimensional manifold lying on a higher-dimensional Euclidean space; these will require different techniques. 

At a high-level, our strategy is to specify a flexible, possibly nonparametric prior over the distribution $q(x)$ and thus implicitly over $p(x)$. Placing a prior over the unconstrained distribution $q(x)$ allows us to avail of standard Bayesian modeling tools. We treat the observations $X$ as the outcome of a rejection sampling algorithm where we propose from $q(x)$ and discard samples falling outside of $\sub$.
Given the observations $X$, {we sample from the posterior distribution over $q(x)$}, implicit in which once more, is all information contained in the posterior over $p(x)$. While $q(x)$ is a simpler object than $p(x)$, its conditional distribution given the observations $X$ is still not easy to sample from.  
In order to do this, we recognize that if we augment the observations $X$ with the rejected proposals from $q(x)$, the conditional inference over $q(x)$ is straightforward. In particular, the rejected proposals (call these $Y$) together with the observations $X$ form i.i.d.\ samples from the unconstrained model $q$. This allows the use of standard MCMC methods for posterior inference over $q$ and imputing $Y$ eliminates any intractable integrals arising from the constraint set $\sub$.

The question now is how to impute the rejected proposals $Y$. In~\citet{rao2016data}, it was shown that the set of rejected samples preceding each observation are exchangeable across different observations. Consequently, in order to reconstruct the rejected samples for any observation $x$, one merely has to sample a new observation from the rejection sampler on $\sub$ and associate all rejected samples to $x$. Concretely, this involves simulating from the proposal distribution $q(x)$ until an acceptance and, after discarding the accepted sample, assigning the remainder to observation $x$. This idea was first proposed by~\cite{beskos06} in the specific setting of 
parameter inference for stochastic differential equations. Repeating this for each observation in the dataset $X$ allows all discarded samples to be imputed.

To describe the overall MCMC algorithm, we parametrize both the constrained density of interest as well as the proposal density by $\param$, writing them as $q(x|\param)$ and $p(x|\param) \propto \mathbbm{1}_\sub(x) q(x|\param)$. As we describe later, $\param$ will represent the parameters of a mixture model {which includes the parameters of the individual components and the component weights}. 
We place a prior $p(\param)$ on $\param$ and given observations $X$, are interested in the posterior given $X$, $p(\param|X)$. To sample from this distribution, we simulate from the distribution $p(\theta,Y|X)$ which has $p(\theta|X)$ as its marginal distribution. 
We sample on this augmented space by repeating two Gibbs steps: 1) simulate the rejected proposals $Y$ given the parameter $\theta$ and 2) update the parameter $\param$ given the rejected samples $Y$, targeting the density $q(\param|X,Y)$. Algorithm~\ref{alg:data_aug} provides the general data augmentation scheme. 
A proof of its correctness can be found~\citet{rao2016data}. 
\\
\begin{algorithm}[ht]
 \SetAlgoLined
 \KwData{The observations $X=\{x_1,\ldots,x_n\}$, and the current parameter values $\param$}
 \KwResult{New parameter value $\tilde{\param}$}
 \For{each observation $x_i$} {
 \While{an accepted sample $\hat{x}$ has not been drawn}{
  draw $y$ independently from $q(\cdot|\param)$\;
 }
 Discard $\hat{x}$ and treat the preceding rejected proposals as ${Y}_i$\;
 }
 Gather all the rejected samples, calling them ${Y}$: 
 ${Y} =  \bigcup\limits_{i=1}^{n} {Y}_{i}$\;
 Update $\param$ from $q(\param|X, {Y})\propto q(X,{Y}|\param) p(\param)$ using 
 any MCMC kernel, calling the new value $\tilde{\param}$\;
 Discard the rejected samples ${Y}$
 \caption{An iteration of MCMC for posterior inference over $p(\param|X)$~\citep{rao2016data}}
 \label{alg:data_aug}
\end{algorithm}

\section{Mixture modeling on constrained spaces}
\label{sec:tmix}
We now move from the abstract setting of the previous section to the focus of this paper, where the proposal density $q(x|\theta)$ is a mixture model and the parameter $\theta$ represents the mixing proportions $\pi$ as well as the component parameters $\beta$. In Bayesian settings, it is typical to place a Dirichlet prior over $\pi$ and when possible, a conjugate prior over $\beta$. Thus, when working with a mixture of Gaussians, $\beta$ would include the mean $\mu$ and covariance $\Sigma$ of the Gaussian components and we would place a Normal-Inverse-Wishart (\text{NIW}) prior over the $(\mu, \Sigma)$ pairs. In nonparametric settings, $\pi$ {lies on the} infinite-dimensional simplex and one might place a stochastic process prior (e.g.\ a Dirichlet process prior,~\citet{Fer1973}) over $\pi$ via e.g.\ a stick-breaking construction~\citep{ishwaran2001gibbs}. Let $\beta_0$ be the hyperparameter for the prior over the component parameters and $\alpha_0$ be the hyperparameter for the distribution over weights. 
We write the latter as a $\text{Dir}$, though it could be either the Dirichlet or a stick-breaking prior.
Then, for the case of Gaussian likelihoods, the prior over parameters can be written as follows: 
\begin{align}
  {\pi} | \alpha_0 \sim \text{Dir}(\alpha_0), \quad 
  (\mu_k,\Sigma_k) | \beta_0 &\sim \text{NIW}(\beta_0),\ \ k = 1,2,\cdots.
  \label{eq:param_prior}
\end{align}

\subsection{Truncated Mixture of Gaussians (TMoG)}
\begin{figure}[ht] 
    \centering 
    \begin{minipage}{0.22\textwidth}
        \includegraphics[width=\textwidth]{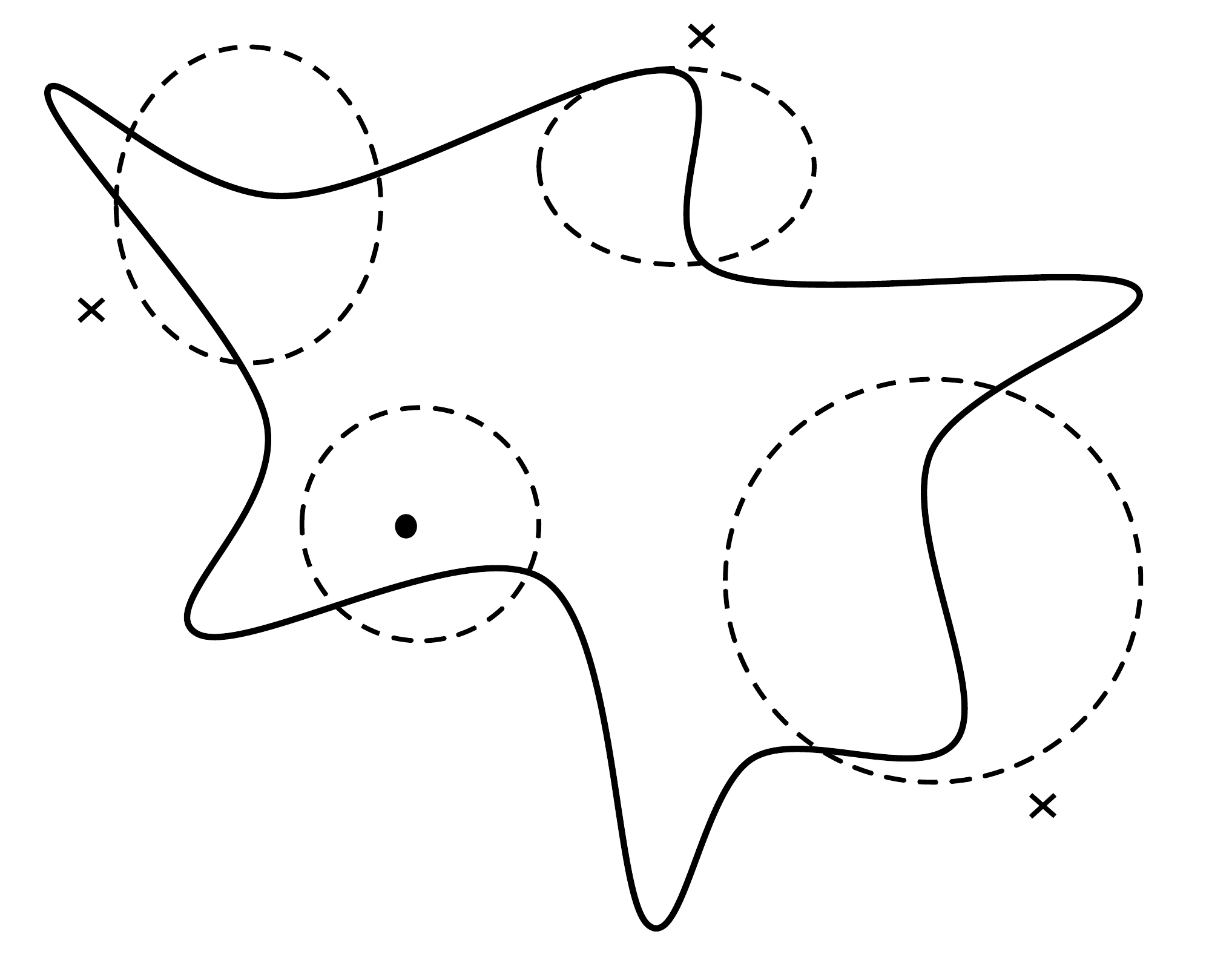}
    \end{minipage}
    \begin{minipage}{0.22\textwidth}
        \includegraphics[width=\textwidth]{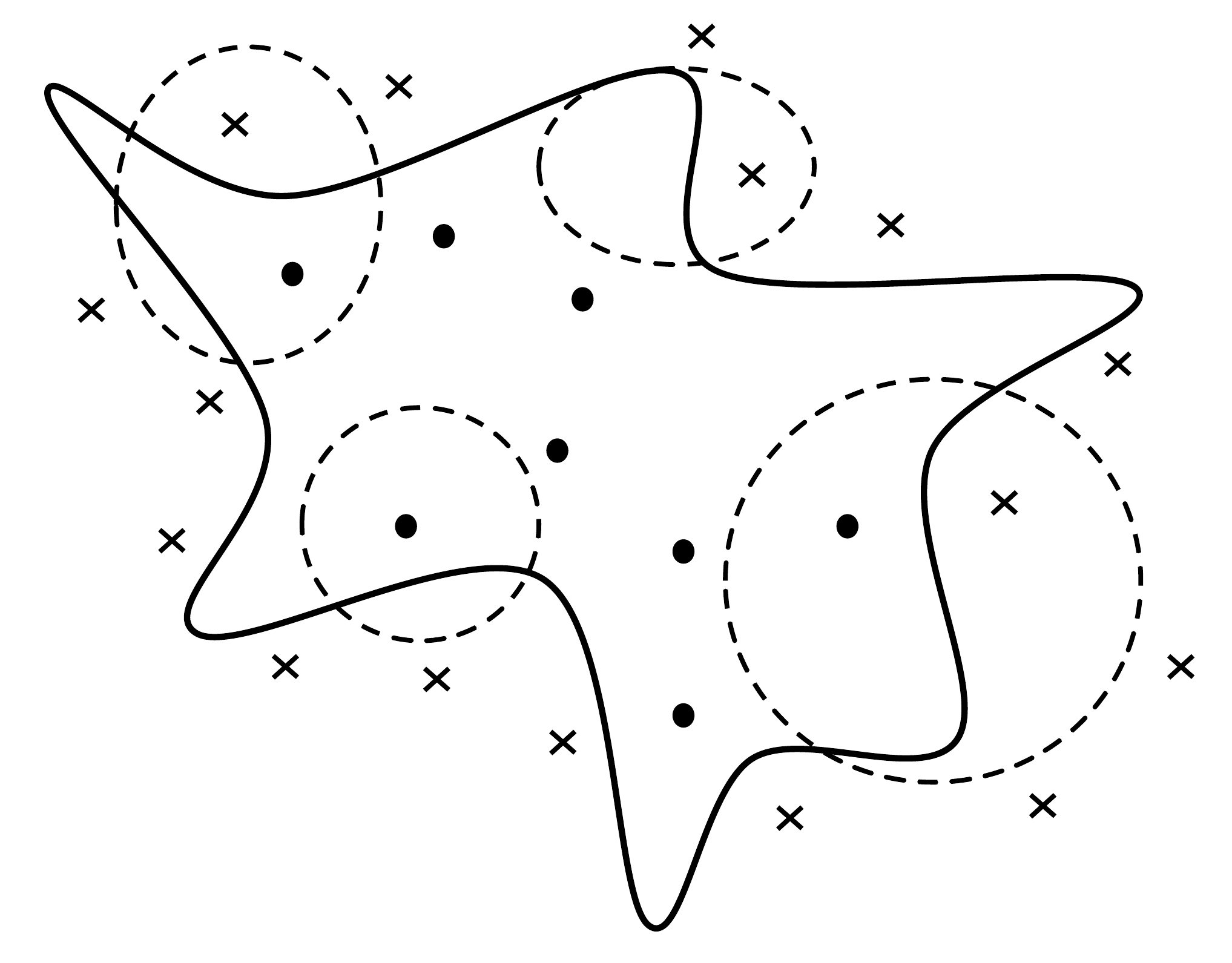}
    \end{minipage}
    \begin{minipage}{0.54\textwidth}
    \caption{The generative process of TMoG starts by selecting a cluster and sampling a single draw from it. 
      If this falls outside $\sub$ (the solid line), again select a cluster and sample another draw from it. 
   The process is repeated until a cluster component produces an accepted draw (left). When the desired number of accepted draws is reached, the set of accepted draws is distributed as TMoG (right). Dashed lines represent cluster likelihoods, dots and crosses are accepted and rejected proposals.}
    \label{fig:gen_proc_tmogt}
    \end{minipage}
\end{figure}

In the first approach we consider, our proposal distribution $q(x|\theta)$ 
is a simple unconstrained mixture model; for concreteness, we use a 
mixture of multivariate Gaussian distributions. Assume $K$ components, 
each with its own mean $\mu_k \in \mathbb{R}^d$ and covariance matrix 
$\Sigma_k \in \mathbb{R}^{d\times d}$. Then the proposal has the form 
\begin{align}
q(x|\theta) = \sum_{k=1}^K \pi_k \mathcal{N}(x|\mu_k, \Sigma_k).
\label{eq:tmog_prop}
\end{align}
As stated, $\theta$ consists of the mixing distribution $\pi$ 
and the $K$ component parameters $\{(\mu_1, \Sigma_1),\ldots,(\mu_K,\Sigma_K)\}$.
In nonparametric settings with a Dirichlet process prior, $K$ is infinity and 
the proposal distribution becomes a Dirichlet process mixture 
of Gaussians~\citep{lo1984class,EscWes1995, rasmussen00}.
We assume our constraint set $\sub$ is some subset of 
$\mathbb{R}^d$ with nonzero Lebesgue measure so that the 
observations follow a density equal to $q(x|\theta)$, truncated to $\sub$ and 
renormalized. We call this distribution a truncated mixture of Gaussians (TMoG): 
\begin{align}
  p(x|\theta) = \frac{\mathbbm{1}_{\sub}(x)\sum_{k=1}^K \pi_k \mathcal{N}(x|\mu_k, \Sigma_k)}
  {\int_{\sub} \mathrm{d}x \sum_{k=1}^K \pi_k \mathcal{N}(x|\mu_k, \Sigma_k)}.
\end{align}
To simulate observations from this model, we make proposals from equation 
\eqref{eq:tmog_prop} until one lies in $\sub$.
We note that since proposals are made independently from the mixture 
distribution of equation~\eqref{eq:tmog_prop}, rejected proposals preceding the same observation 
need not belong to the same cluster. Figure \ref{fig:gen_proc_tmogt} 
describes the generative process of this model in more detail. 

Given observations $X$, MCMC inference over the parameters involves first imputing the rejected proposals conditioned on the parameters and then updating the parameters given these imputed variables. Having updated the parameters, we discard the rejected samples and repeat the process. We describe the steps of the overall Gibbs sampler 
below. 

\begin{description}
  \item[Imputing the rejected proposals $Y$:] 
To impute the rejected proposals given $X$ and $\theta = (\pi, \mu, \Sigma)$, we follow 
steps 1 to 6 of Algorithm~\ref{alg:data_aug}, proposing from the mixture of normals 
(equation~\eqref{eq:tmog_prop}) to generate a pseudo-dataset of the same 
size as $X$ and keeping all the rejected proposals generated along the way. 
Call these $Y$ and write $R$ for the total number of elements in $Y$.
Each element $y_i$ lies outside the constraint set $\sub$ and is {associated with a mixture component $c^*_i$ from which it was drawn.} 
Write $C^*=\{c^*_1,\ldots,,c^*_{R}\}$ for the set of cluster assignments of the imputed proposals.
$Y$ and $C^*$, together {with} $X$ will be used to 
update the parameters $\theta$ as well as the cluster assignments of 
the observations (write these as ${C} =  \{c_1,\ldots, c_n\}$).
The joint probability is 
\begin{align}
p({X}, {Y}, {\pi}, {\mu}, \Sigma, {C}, {C^*} | \alpha_0, \beta_0)
=& \text{Dir}(\pi | {\alpha}_0)\prod_{k=1}^K\text{NIW}({\mu_k, \Sigma_k} |\beta_0) 
\prod_{i=1}^N \pi_{c_i} \mathcal{N}(x_i |\mu_{c_i},\Sigma_{c_i}) 
\prod_{r=1}^{R} \pi_{c_r^*} \mathcal{N}(y_{r}|\mu_{c_r^*}, \Sigma_{c_r^*}).
\label{equation:jointlike}
\end{align}
Updating $C, \pi$ and the $(\mu_k,\Sigma_k)$'s is now straightforward as described next.
  \item[Updating the mixing proportions $\pi$:] 
Write $n_k$ and $m_k$ for the total number of observations and rejected 
samples (respectively) in cluster $k$. These are easily calculated from $C$ and $C^*$. 
For a Dirichlet distribution prior over $\pi$, the Gibbs conditional over $\pi$ takes the simple form
\begin{align}
p({\pi}|{X}, {Y}, {\theta}, {C}, {C^*}, \alpha_0, \beta_0) 
&= \text{Dir}(n_1 + m_1 + \alpha_0, \ldots, n_K + m_K +\alpha_0).
\end{align}
For a nonparametric stick-breaking prior over $\pi$, the conditional update for $\pi$ is a simple adaptation of standard methodology (such as in~\cite{ishwaran2001gibbs}). The key point implied by equation~\eqref{equation:jointlike} for any prior over $\pi$ is to include {\em both} observations and rejected samples in the cluster counts.
  \item[Updating the cluster assignments $c_i$:] 
    Use $\neg i$ to represent quantities calculated after excluding 
    observation $i$. The conditional distribution for $c_i$, the 
    cluster assignment of observation $i$, is then given by
\begin{align}
  p(c_i = k | {C}^{\neg i}, {C^*}, {X}, {Y}, {\theta}, \alpha_0, \beta_0) &\propto \pi_{c_i} \mathcal{N}(x_i|\mu_{c_i}, \Sigma_{c_i}).
\end{align}
Conditioned on $\pi$, all $c_i$'s can be updated independently. 
One can also marginalize out $\pi$ and update $c_i$'s sequentially. 
Now, {if} the $c_i$'s follow a P\'olya urn/Chinese restaurant process update~\citep{neal2000markov}, where again, we must consider {\em both} the number of observations as well as the number of rejected proposals at any cluster:
\begin{align}
p(c_i = k | {C}^{\neg i}, {C^*}, {X}, {Y}, {\theta}, \alpha_0, \beta_0) &\propto 
(n_k^{\neg i}+m_k^{\neg i}) \mathcal{N}(x_i|\mu_{c_i}, \Sigma_{c_i}).
\end{align}
For a new cluster, we replace $(n_k^{\neg i}+m_k^{\neg i})$ with the 
concentration parameter $\alpha_0$.

\item[Updating the cluster parameters $(\mu_k,\Sigma_k)$:] 
The conditional distribution for the cluster parameters depends both on observations and rejected 
samples assigned to that cluster. Writing $C_k$ and $C^*_k$ for the indices of observations and rejected samples assigned to cluster $k$ and $\theta^{\neg k}$ for all parameters except $(\mu_k,\Sigma_k)$, we 
have
\begin{align}
p(\mu_k, \Sigma_k | {\theta}^{\neg k}, {X}, {Y}, {C}, \alpha_0, \beta_0) 
&\propto p(\mu_k, \Sigma_k|\beta_0)\prod_{i \in C_k} p(x_{i}|\mu_k, \Sigma_k) 
\prod_{r\in C^*_{k}} p(y_{r}|\mu_k, \Sigma_k).
\end{align}
The parameters for all components can be updated independently and with 
a conjugate Normal-Inverse-Wishart prior, this distribution is easy to sample from.
\end{description}

\subsection{Mixtures of Truncated Gaussians (MoTG)}
\label{sec:mtog}
\begin{figure}[ht]
    \centering
    \begin{minipage}{0.22\textwidth}
        \includegraphics[width=\textwidth]{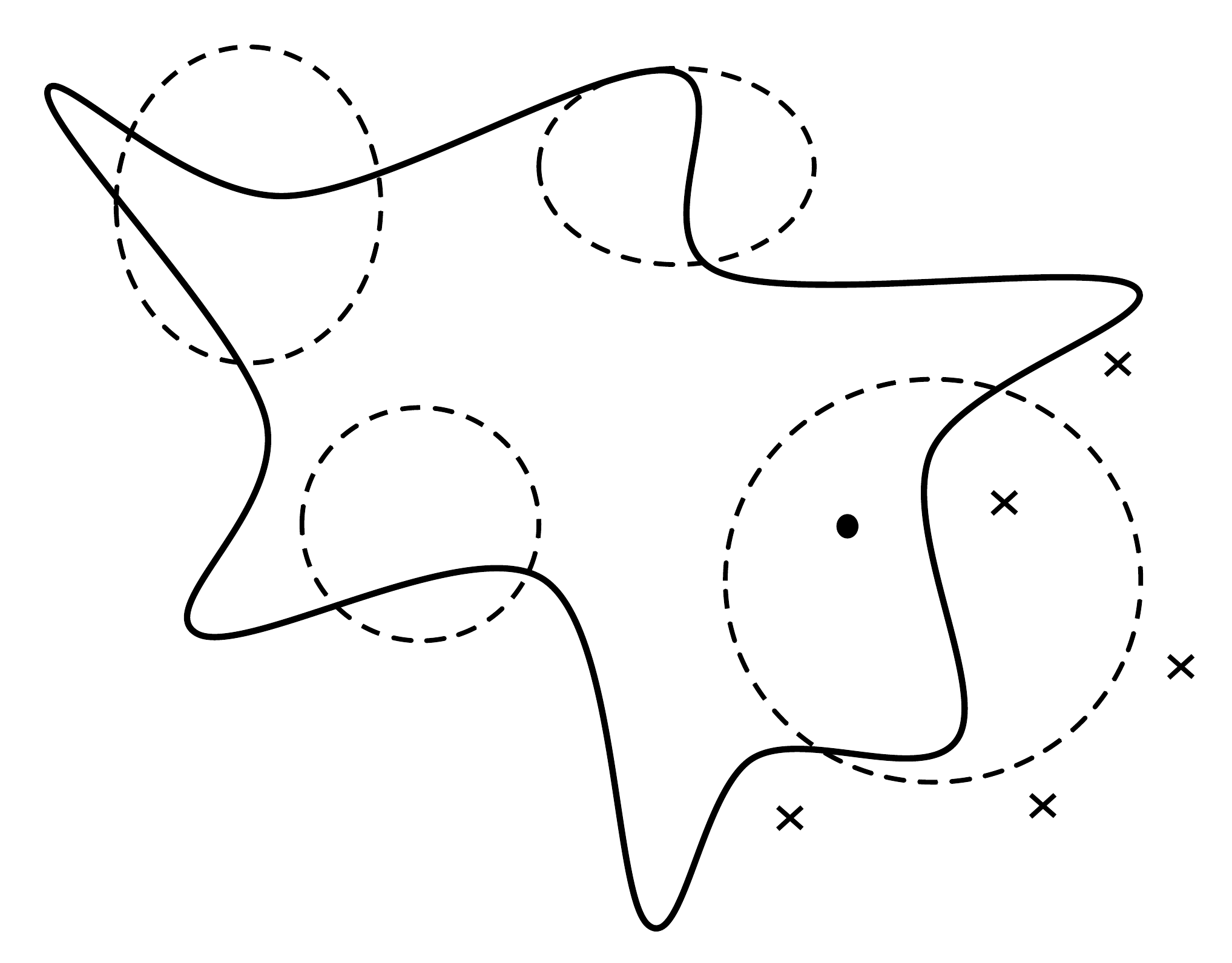}
    \end{minipage}
    \begin{minipage}{0.22\textwidth}
        \includegraphics[width=\textwidth]{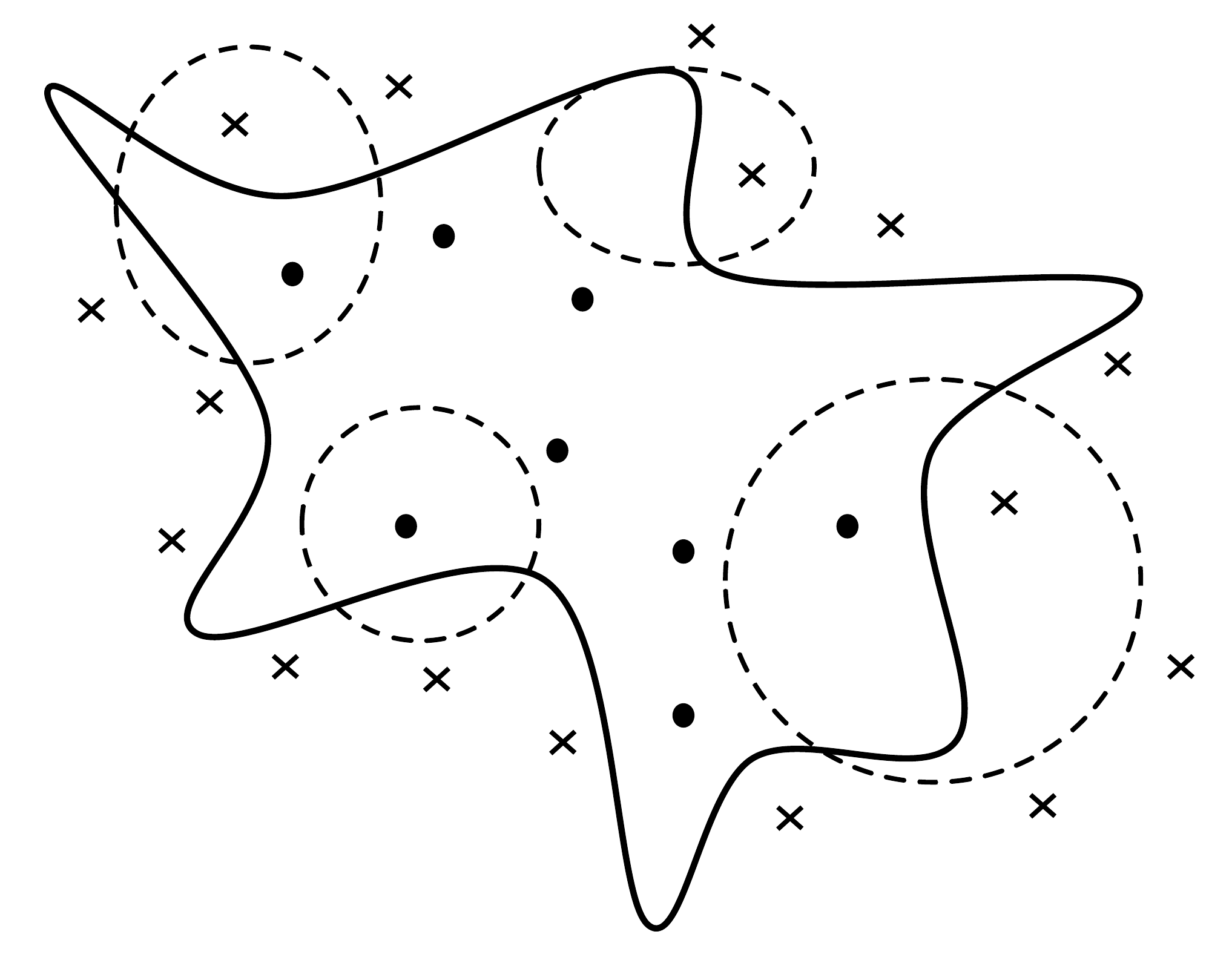}
    \end{minipage}
    \begin{minipage}{0.54\textwidth}
    \caption{The generative process of MoTG starts by picking a cluster and sampling from the associated unconstrained distribution until an acceptance (left). This is repeated until the desired number of acceptances are made (right). The set of all accepted draws are distributed as MoTG on $\sub$. The solid line is the constraint set $\sub$, the dashed lines represents the cluster components, dots and crosses are accepted and rejected proposals. }
    \label{fig:gen_proc_motg}
    \end{minipage}
\end{figure}

In the previous section, we modeled an unknown density on a constrained 
space with a truncated mixture model (in particular, a truncated mixture 
of Gaussians). In this section, we take a second approach, 
modeling the density as a {\em mixture of truncated distributions} 
(for concreteness, a mixture of truncated Gaussians). 
For a subset $\sub$ of a $d$-dimensional Euclidean space, write 
$\tn(x|\mu,\Sigma)$ for a Gaussian with mean $\mu$ and covariance $\Sigma$ 
restricted to that subset:
\begin{align}
  \tn(x|\mu,\Sigma) &= 
  \frac{\mathbbm{1}_{\sub}(x)\mathcal{N}(x|\mu,\Sigma)}{\int_\sub\mathcal{N}(x|\mu,\Sigma)dx}.
\end{align}
A mixture of $K$ truncated Gaussians, with parameters 
$\{(\mu_1,\Sigma_1),\ldots,(\mu_K,\Sigma_K)\}$ and with mixing proportion 
$\pi$ has probability density given by: 
\begin{align} 
p(x|\theta) & = \sum_{k=1}^K \pi_k \tn(x|\mu_k, \Sigma_k).
\end{align}
Unlike the truncated mixture of Gaussians which involves a single intractable 
normalization constant (equation~\eqref{eq:tmog_prop}), the equations
above show that the mixture of truncated Gaussian involves $K$ (albeit 
simpler) intractable normalization constants.
As before, we place Dirichlet/stick-breaking priors on $\pi$ and 
a conjugate Normal-Inverse-Wishart prior on the components parameters 
$(\mu_k,\Sigma_k)$.
The generative process then follows: 
\begin{align*}
\pi | \alpha_0 &\sim \text{Dir}(\alpha_0), \quad 
(\mu_k, \Sigma_k)| \beta_0 \sim \text{NIW}(\beta_0), \ \ 
  k=1,\ldots,K, \\
  X_i | c_i, \{(\mu_k, \Sigma_k)\}_{k = 1}^K &\sim \tn(x_i|\mu_{c_i}, \Sigma_{c_i}), \quad
c_i | \pi \sim \text{Discrete}(\pi), \ \ i=1,\ldots,N.
\end{align*}
Having chosen the cluster $c_i$ of observation $i$ from the distribution 
$\pi$, the challenge now is to sample from the corresponding truncated 
normal distribution $\tn(x|\mu_{c_i},\Sigma_{c_i})$. To do this, we 
again use rejection sampling, now proposing from the unconstrained Gaussian distribution 
$\mathcal{N}(x|\mu_{c_i},\Sigma_{c_i})$ until acceptance. Note that now, 
unlike with the truncated mixture of Gaussians, all rejected samples 
associated with an observation come from the same component as that 
observation. Figure \ref{fig:gen_proc_motg} outlines this process in more detail. 
This results in subtle differences in the associated MCMC 
sampler where now, rejected proposals of each observation must be imputed 
in a cluster-specific manner. Having imputed these, we must update 
cluster assignments and parameters, again in a manner slightly different from the TMoG case. As before, at the 
end of these updates, we discard the imputed samples and repeat. We describe the full Gibbs sampling procedure below. 


\begin{description}
  \item[Imputing the rejected proposals $Y$:] 
Unlike TMoG, where each observation is an accepted proposal from a mixture of Gaussians, now each observation is an accepted proposal from the single Gaussian component to which it was initially assigned. Accordingly, to impute the rejected proposals 
for observation $x_i$ belonging to cluster $c_i$, we repeatedly propose from component $c_i$ until acceptance and keep all the rejected proposals. We now have to keep track of which rejected proposals belong to which observation and write $Y_i$ for the set of rejected samples preceding observation $x_i$ and $R_i$ 
for the cardinality of $Y_i$. The joint probability density is then 
\begin{align}
p(X, Y, \pi, \theta, c | \alpha_0, \beta_0) 
= &\text{Dir}(\pi | \alpha_0) \prod_{k=1}^K \text{NIW}(\mu_k,\Sigma_k|\beta_0) 
    \prod_{i=1}^N  p(c_i|\pi) \left(
  \mathcal{N}(x_i|\mu_{c_i}, \Sigma_{c_i}) 
\prod_{r=1}^{R_i} \mathcal{N}(y_{i,r}|\mu_{c_i}, \Sigma_{c_i}) \right).
\label{equation:jointlike_motg}
\end{align}
We use this to update the latent variables as described below.
  \item[Updating the mixing proportion $\pi$:] 
Write $n_k$ for the number of observations assigned to component $k$.
The conditional distribution for $\pi$ follows a Dirichlet distribution:
\begin{align}
p(\pi|X, Y, \theta, C, \alpha_0, \beta_0) &\propto 
\text{Dir}(n_1 + \alpha_0, \ldots, n_K + \alpha_0).
\end{align}
Unlike TMoG, this distribution does not involve the 
rejected samples at all. This is because for each observation, a cluster 
is chosen from $\pi$ only once, with all rejected proposals assigned to that
same component. For any other prior over $\pi$ (e.g.\ a Dirichlet process),
the conditional update rule remains unchanged from standard methodology, involving only 
cluster assignment counts of observations.
  \item[Updating the cluster assignment $c_i$:] 
Unlike TMoG, updating $c_i$ now involves the rejected 
proposals asssociated with observation $i$ and has conditional 
distribution
\begin{align}
p(c_i = k | {C}^{\neg i}, X, {Y}, \theta, \alpha_0, \beta_0) 
&\propto \pi_{k} \mathcal{N}(x_i|\mu_{k}, \Sigma_k) 
\prod_{r=1}^{R_i}\mathcal{N}(y_{ir}|\mu_{k}, \Sigma_k).
\end{align}
This is a consequence of the fact that the cluster assignment is made only 
once, after which proposals are made from that cluster until acceptance.
This also accounts for the fact that without the rejected proposals, this 
distribution would be proportional to $\pi_k \tn(x_i|\mu_k,\Sigma_k)$ 
and would involve the intractable normalization constant of component $k$. 
Imputing the rejected proposals avoids having to calculate this quantity, though 
now, the associated rejected proposals are transferred along with that 
observation to a new cluster. If $\pi$ were marginalized out (e.g.\ under 
a Chinese restaurant process), the cluster update rule becomes
\begin{align}
  p(c_i = k | {C}^{\neg i}, {C^{\neg i *}}, {X}, {Y}, {\theta}, \alpha_0, \beta_0) &\propto 
n_k^{\neg i} \mathcal{N}(x_i|\mu_{k}, \Sigma_{k})
\prod_{r=1}^{R_i}\mathcal{N}(y_{ir}|\mu_{k}, \Sigma_k).
\end{align}
Here $C^{\neg i *}$ refers to all rejected proposals except those associated with $i$.
\item[Updating the cluster parameters $(\mu_k,\Sigma_k)$:] 
For the same reason as above, updating cluster parameters also 
require conditioning on the imputed samples. The conditional 
distribution is identical to that for TMoG: 
\begin{align}
p(\mu_k, \Sigma_k | {\theta}^{\neg k}, {X}, {Y}, {C}, \alpha_0, \beta_0) 
&\propto \text{NIW}(\mu_k, \Sigma_k|\beta_0)\prod_{i \in C_k} 
 \mathcal{N}(x_{i}|\mu_k,\Sigma_k) 
\prod_{y\in Y_{i}} \mathcal{N}(y|\mu_k,\Sigma_k). \nonumber
\end{align}
With a Normal-Inverse-Wishart prior for the Gaussian mixture 
model, this distribution is easy to sample from.
\end{description}

\section{Extensions to improve MCMC mixing }
\label{sec:thrdars} 
The MCMC schemes for the two models outlined earlier target the exact posterior over the distribution $q$ and through this, the posterior distribution over $p$. 
In our experiments, we see that despite their exactness, these MCMC algorithms comes at the cost of a high variance, especially in higher-dimensional settings. At a high level, this can be attributed to the nonparametric prior over $q$ being too flexible and placing too much probability on proposal distributions $q$ that themselves place significant probability outside of $\sub$. The observations $X$ govern how $q$ assigns probability within $\sub$ characterizing the distribution $p$ that we care about. Right outside the boundary of $\sub$, the observations indirectly constrain $q$ because of prior smoothness assumptions. As we move away from $\sub$, the influence of the observations starts to wane, and the posterior over $q$ reverts back to the prior.
In this part of $\euc$-space, the MCMC sampler explores what is effectively the prior over $q$ by simulating rejected proposals $Y$ from $q$ and then simulating $q$ from its posterior distribution given these rejections.
The resulting coupling between $Y$ and $q$ can cause the MCMC chain to mix very poorly, and the estimate of the distribution over $q$ can have large variance. We reiterate that we do not actually care about $q$ outside $\sub$, rather, this only serves to avoid underestimating density and smoothness estimates at the boundary.

To understand the mechanics of this more clearly for mixture models, observe that as the MCMC algorithm explores the parameter space, a cluster will occasionally be produced under which the subset $\sub$ has low probability. 
Producing an accepted proposal from this component can require a large number of rejected proposals. 
As a consequence, the MCMC algorithm will experience a slow down, taking a long time to move through an MCMC iteration, with the increased number of 
rejected proposals also increasing coupling between MCMC iterations. 
The large number of rejected samples will start to swamp out the observations $X$ causing the cluster parameters to drift away from the observations and $\sub$, further exacerbating the issue. The net effect is longer computation times and larger variances in the MCMC estimates. Since MoTG proposes from a particular Gaussian component until acceptance (unlike TMoG, which picks a new component for each proposal), we expect that this effect is worse for MoTG than TMoG. 

%
An obvious remedy is to choose the hyperparameters of the nonparametric prior to concentrate on $q$'s with high $q(\sub)$. 
Setting parameters in this manner is not easy however, especially with complex, asymmetric constraint sets such as those shown in Figure~\ref{fig:examples}.
%
We instead take a more direct approach,
introducing a single additional parameter $\rho \in (0,1)$, corresponding {to} the belief that $q$ is restricted to distributions satisfying 
$q(\sub) = \rho$. 
Small values of $\rho$ imply that $q$ assigns relatively large probability outside $\sub$ resulting in a large number of rejected samples. This is useful when the constraint set $\sub$ is complex and one expects significant probability mass at the edges. 
Values close to one imply beliefs that the proposal distribution need not be too complex outside. Note that since we use a nonparametric prior over $q$, we do not have to be too careful about $\rho$. In fact, we expect that for any setting of $\rho$, the posterior is asymptotically consistent provided a reasonable prior is placed over the variance of the mixture components (see e.g.\ ~\citet{canale2017}). $\rho$ however can have a significant impact on mixing and finite-sample performance. From our experiments, we recommend $\rho = 0.5$ as a reasonable default.


Write $\mathcal{Q}_\rho$ for the space of probability measures that assign probability $\rho$ to $\sub$.
Write $\text{TMoG}_\rho$ and $\text{MoTG}_\rho$ for TMoG and MoTG prior restricted to this set. Working with these as priors over $q$ is closely related to the approach outlined in~\citet{margBNP} where the authors are interested in nonparametric Bayesian priors given specifications of certain marginal distributions. 
There the posterior simulation was approximate, involving kernel density estimates of marginal probabilities under the original prior. Unlike that work which considers general marginal specifications, we have a simple and specific requirement: any $q$ simulated from the nonparametric prior (whether TMoG or MoTG) must satisfy $q(\sub) = \rho$. 
Further, we only introduce this constraint for computational reasons and can tolerate slight violations (which are normalized out in $p(x) \propto \mathbbm{1}_\sub(x)q(x)$). We thus propose a much simpler approximation strategy that becomes more accurate as the number of observations increases. 
As before, the Gibbs sampler involves three steps:
\begin{description}
  \item[Imputing the rejected proposals $Y$:] 
The fact {that} $q(\sub) = \rho$ means that the number of rejected samples is geometrically distributed with success probability 
{$\rho$}.
This, and the fact that they are exchangeable~\citep{rao2016data} allows a simple simulation approach: first sample the number of rejected samples from the geometric distribution with parameter 
{$\rho$}, and then use rejection sampling to simulate their locations outside $\sub$.
  \item[Updating cluster parameters $\beta$ and weights $w$:] 
  This step presents a challenge since we need to update these while satisfying the constraint $q(\sub) = \rho$.
  However, we observe that for reasonably large datasets, we can drop this constraint without introducing too much error. In particular, with $n$ observations, our imputation scheme introduces $\frac{1-\rho}{\rho}n$ rejected samples. For large $n$, standard consistency results for DP mixture models tell us that the posterior will concentrate around $q$'s that satisfy $q(\sub) = \rho$. Since we are really interested in $p(x) \propto \mathbbm{1}_{\sub}(x)q(x)$, any deviation of $q(\sub)$ from $\rho$ is renormalized out and introduces minimal error into our final analyses. Our experiments confirm this.
  \item[Updating cluster assignments $C$:] 
    Having instantiated the proposal distribution $q$ through its parameters $\theta = (\beta, w)$, updating the cluster parameters is identical to corresponding steps in the original TMoG and MoTG samplers. 
\end{description}

\subsection{Additional simplifications}
In the original MCMC samplers, to impute rejected proposals for an observation, proposals are made until one is accepted. In the previous section, restricting ourselves to proposal distributions satisfying $q(\sub)=\rho$, we simulate a geometrically distributed number of rejections. 
For a dataset with $n$ observations, the average number of rejected proposals is $n\frac{1-\rho}{\rho}$. 
For $n$ not to small, we can 
approximate the number of rejected samples with its average, avoiding the need for any stochastic simulation and simplifying the algorithm.

Secondly, in many settings, the practitioner might be conservative choosing $\rho$, so that the number of rejections from the geometric distribution {\em exceeds} that produced by the original algorithm. To avoid this undersirable inefficiency, we suggest a further simplifications, changing the first step of the Gibbs sampler to: 
\begin{description}
  \item[Imputing the rejected proposals $Y$:] 
simulate rejected proposals from $q$ outside $\sub$ until $n\frac{1-\rho}{\rho}$ 
or $n$ acceptances are produced, whichever occurs first. 
\end{description}
Effectively, we are changing our marginal specification, restricting our nonparametric prior to $q$'s satisfying $q(\sub) \ge \rho$ (instead of the strict equality from before). We repeat that changing the marginal specification in this way does not affect the asymptotic properties of our nonparametric model. For finite number of observations, a small value of $\frac{1-\rho}{\rho}$ simplifies computation and improves MCMC mixing. Making this too small can introduce edge effects as not enough observations are produced outside $\sub$, hurting the ability to {estimate} large density values at the boundaries.

In our experiments, we consider different settings of this threshold, setting $\frac{1-\rho}{\rho}$ to 0.5, 1, 5, and 50. We refer to the final simplified sampler as a  {\em thresholded sampler}. 
Section \ref{sec:simstudy} studies the trade-offs involved for different threshold settings and the performance of the algorithm in terms of time and the predictive likelihood for various simulation studies. Our experiments suggest a threshold of $1$, corresponding to $\rho=.5$.

\section{Simulation studies}
\label{sec:simstudy}
We apply our models and algorithms to a number of synthetic and real datasets with both simple and complex constraints. We study the modeling and computational trade-offs between the truncated mixture of Gaussian (TMoG) and the mixture of truncated Gaussian (MoTG) models as well as between different settings of $\rho$ (or equivalently, different thresholds for the number of rejections). 
For synthetic experiments, we generate datasets with 500 observations, and evaluate performance by using $80\%$ of the data as training and the remaining $20\%$ as held-out test data. 
We evaluate test-likelihood using importance sampling, verifying these numbers with numerical integration when possible. 

We consider 6 different settings of $\rho$, corresponding to threshold values of 0, 0.5, 1, 5, 50, and infinity. Here, 0 means no data augmentation, so that the data are modeled with an unconstrained mixture model that is not cognizant of truncation boundaries. We will see that this can result in inappropriately low probability at the boundaries of the constraint set, as the mixture model tries to explain the absence of any observations outside $\sub$.  A threshold of infinity means no thresholding, corresponding to our original MCMC algorithms. The remaining settings limit the maximum number of rejections per observation, corresponds to progressively smaller lower-bounds on $q(\sub)$. 

We repeat each experiment 100 times, plotting the median and 25\% and 75\% quantiles, for both test log-likelihood and compute time. Our MCMC samplers used a total of 5000 iterations, with the first 2000 discarded as burn-in. In all experiments, we use a stick-breaking prior truncated to 50 components, with the concentration parameter set to 1. 

\subsection{Truncated Gaussian on the unit interval}
We start with a simple one-dimensional setting, where $\sub$ is the unit interval. 
This does not really require our methodology because all normalization constants can be calculated numerically, or since one might choose a different model like a mixture of Beta distributions. 
Nevertheless, this provides a simple test case to exactly evaluate the 
effect of different modeling and algorithmic choices. In higher dimensions (such as the unit square or hypercube), the limitations of standard methods become more evident, and our methodology will be necessary to capture correlation structure as well as high probabilities at the edges. 
We consider a dataset of observations drawn from a truncated Gaussian centered at the edge of the interval, 
$\mathcal{N}(0, 0.05)$,  and model this using TMoG and MoTG, placing on the components a Normal-Inverse-Gamma prior:
$  \mu | (\sigma^2, \mu_0, \lambda_0)  \sim \mathcal{N}(\mu_0, \sigma^2 \lambda_0), \quad 
  \sigma^2 | (\alpha_0, \beta_0) \sim \text{Inv-Gamma}(\alpha_0, \beta_0). 
  $
We set the mean $\mu_0$ to the true mean, $0$ 
and set parameters $\lambda_0, \alpha_0$ and $\beta_0$ to $2, 2$, and $0.05$ 
respectively. 


\begin{figure}
\begin{minipage}{0.65\textwidth}
    \centering
    \includegraphics[width=.98\textwidth]{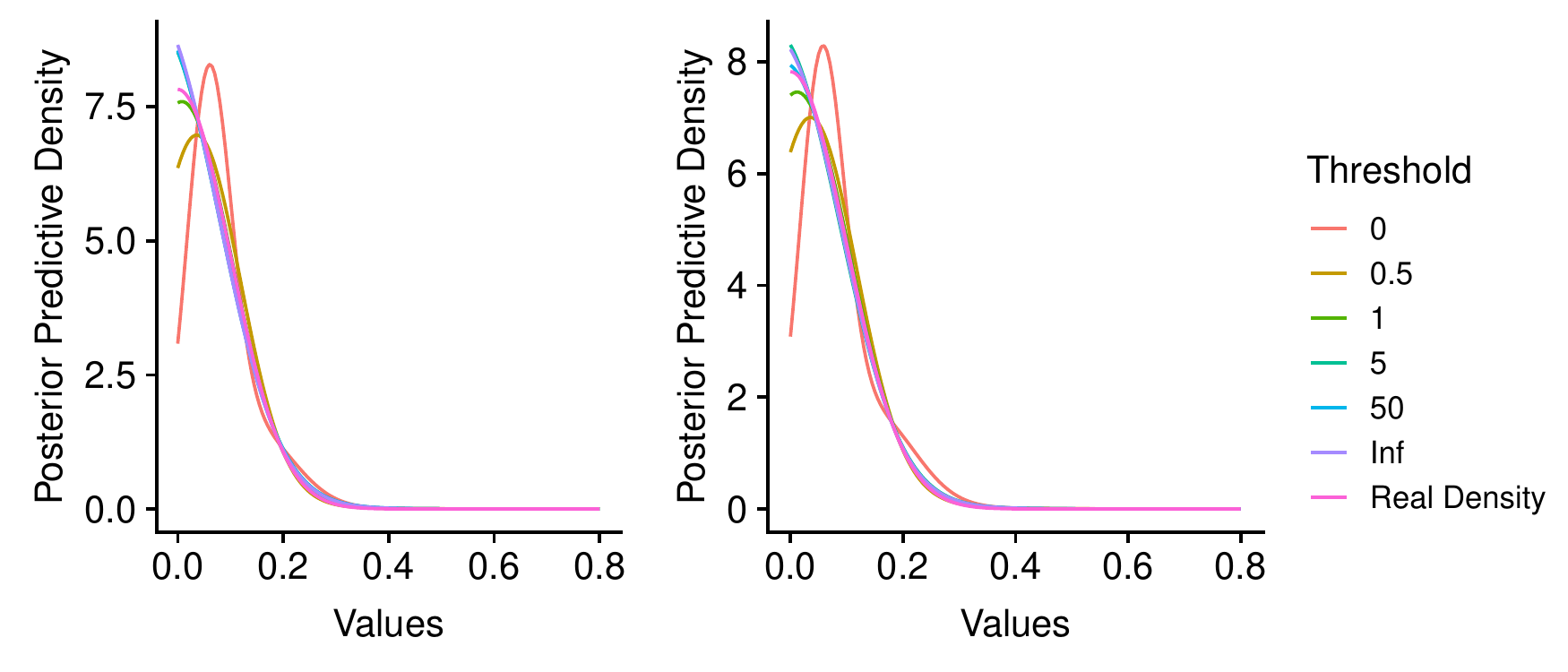}
  \end{minipage}
\begin{minipage}{0.34\textwidth}
    \caption{{Posterior} {predictive} density for MoTG (left) and TMoG (right) for different threshold settings. The settings $0$ and $1$ clearly 
    understimate the density at the origin.}
    \label{fig:postpredgauss_motg_tmog}
  \end{minipage}
\end{figure}
\begin{figure}[ht]
\centering
\includegraphics[width=\textwidth]{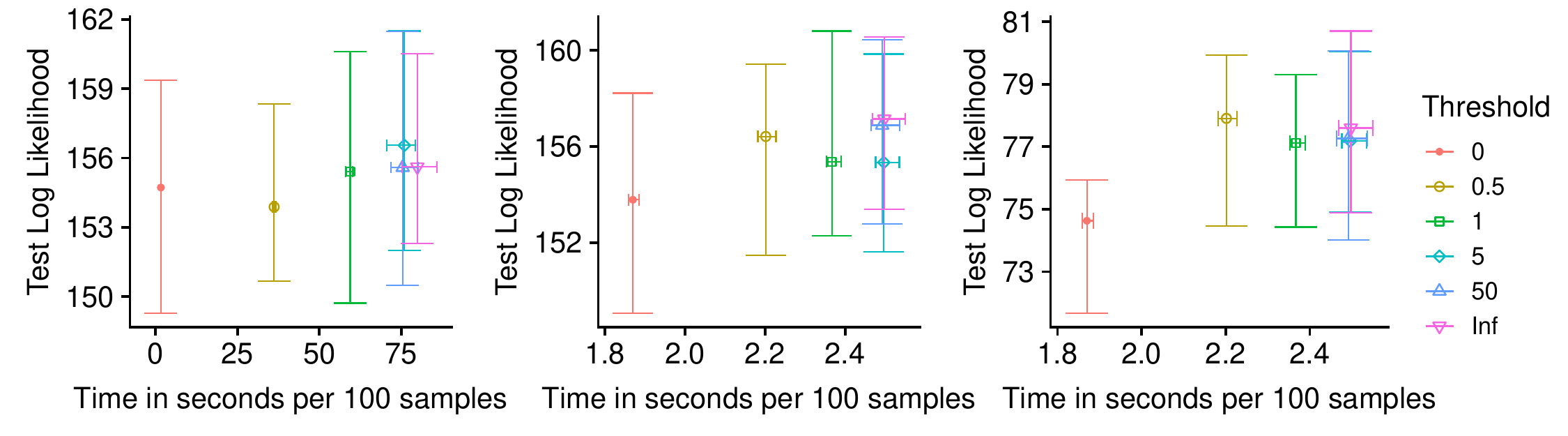}
\caption{Speed-accuracy performance for MoTG (left) and TMoG (center) for $\mathcal{N}(0, 0.1)$. The rightmost panel shows TMoG with test data biased towards the edges.}
\label{fig:gauss_perf_plot}
\end{figure}

\begin{figure}[ht]
    \centering
    \begin{subfigure}[b]{0.45\textwidth}
    \includegraphics[width=\textwidth]{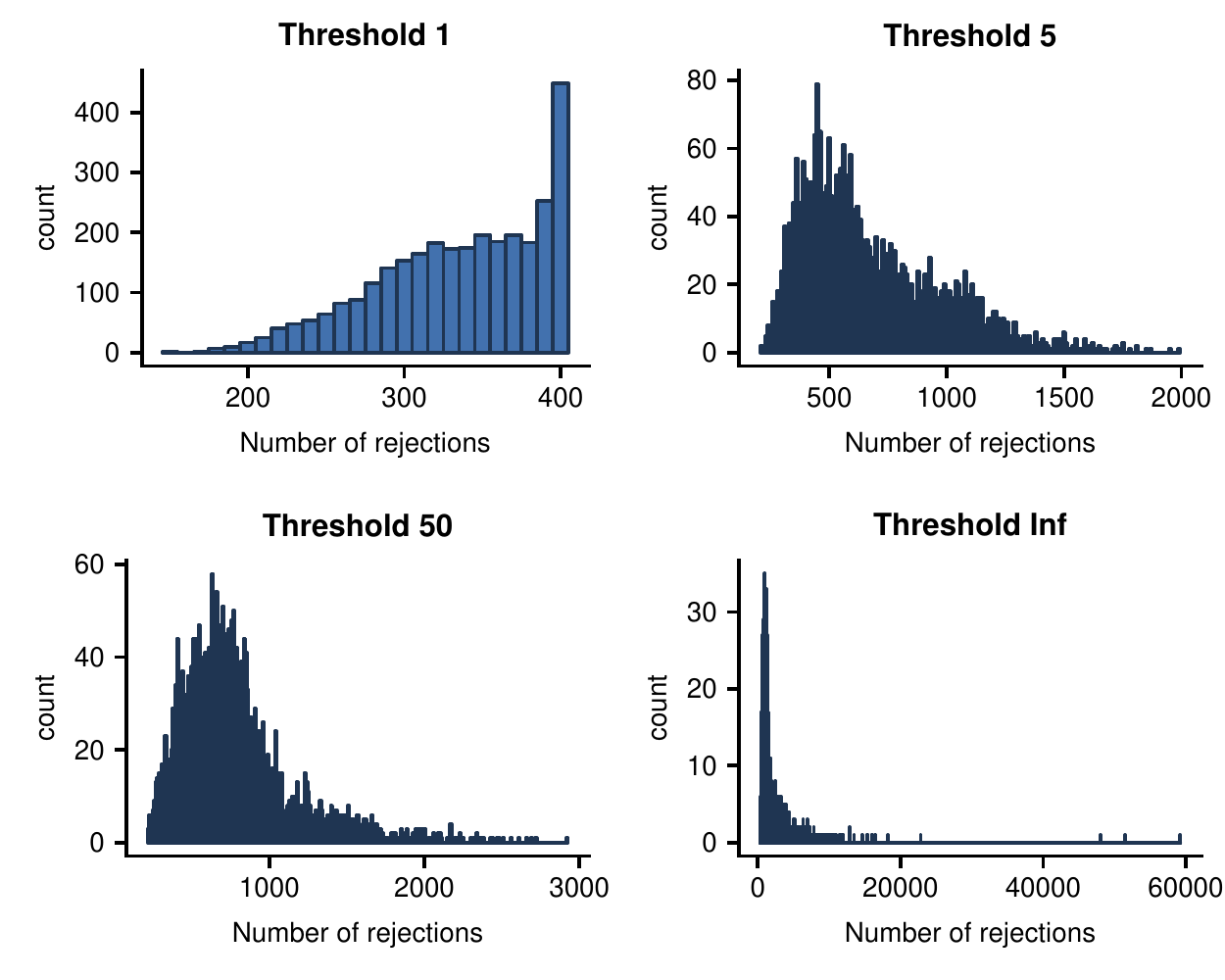}
    \end{subfigure}
    \quad
    \begin{subfigure}[b]{0.45\textwidth}
    \includegraphics[width=\textwidth]{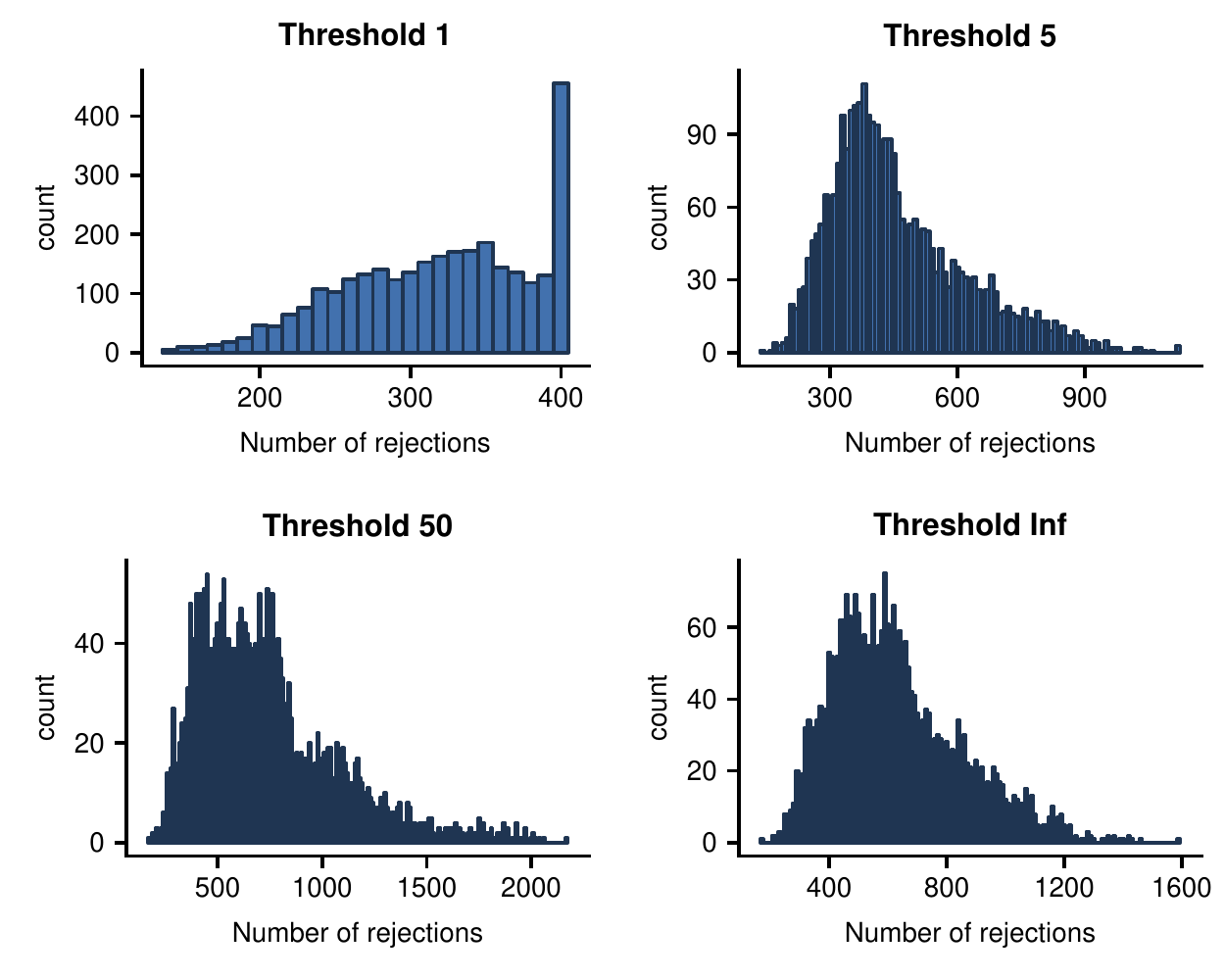}
    \end{subfigure}
    \caption{Histogram of number of rejections for 
      MoTG (left) and TMoG (right) for the truncated $\mathcal{N}(0,0.1)$ 
  distribution}
    \label{fig:histrej_tmogt_motgt}
\end{figure}



Figure \ref{fig:postpredgauss_motg_tmog} shows the {mean} {predictive} density for MoTG (to the left), and TMoG (to the right) for this dataset, for different threshold settings. 
For a threshold equal to 0, the {posterior} experiences a clear shift away from the left boundary; this is despite the fact that we use a flexible mixture of Gaussians to model this distribution.
Essentially, the model struggles to reconcile the abrupt change in the number of observations across the boundary and compromises by smoothing 
across the boundary, resulting in a moderate (rather than high) density at the edge. 
While it might be possible to try to get around this using a prior allowing very peaked components, this will not account for the smoothness of the density inside the interval. 
This emphasizes the importance of accounting for boundary effects in modeling constrained data. 

Figure \ref{fig:gauss_perf_plot} presents the speed-accuracy trade-off for different threshold settings, each plotting the log-likelihood of the test dataset on the y-axis against the run-time for 100 iterations on the x-axis. Results for MoTG are to the left and TMoG to the middle. 
As expected, we see an increase in run-time as the threshold increases, though there are no significant differences for {thresholds} larger than $1$ (indicating that these thresholds are never reached).
Interestingly, the qualitative degradation resulting from a {threshold} of $0$ does not manifest itself quantitatively, with no difference in test performance for different settings. 
This is partly because not enough test observations lie close to the edge to significantly affect the log-likelihood. We will see such effects later with sharper densities and higher-dimensional settings. If however we bias the test dataset to favor observations near the edges, we see a clear drop in performance when the threshold is set to $0$. Here, we selected the test set from observations in the interval $(0,.05)$.

In terms of efficiency, 
we find that TMoG runs significantly faster than MoTG. 
This is a result of the generative process of MoTG where, having picked a cluster, one must repeatedly sample from it until an acceptance.
When the chosen cluster assigns low probability to $\sub$, we get a large 
number of rejections before acceptance. This can also initiate a runaway 
event, with the rejected proposals (that lie outside $\sub$) drawing the 
cluster away from $\sub$ while increasing its mixture selection 
probability, resulting in even more rejections. Figure 
\ref{fig:histrej_tmogt_motgt} demonstrates this by plotting histograms of 
the number of rejected proposals for different threshold settings. When 
this threshold is infinite, the number of rejected samples is more than an order of magnitude larger for MoTG.

\subsection{Beta Distribution}
\begin{figure}[]
\begin{minipage}{0.7\textwidth}
    \centering 
    \includegraphics[width=0.8\textwidth]{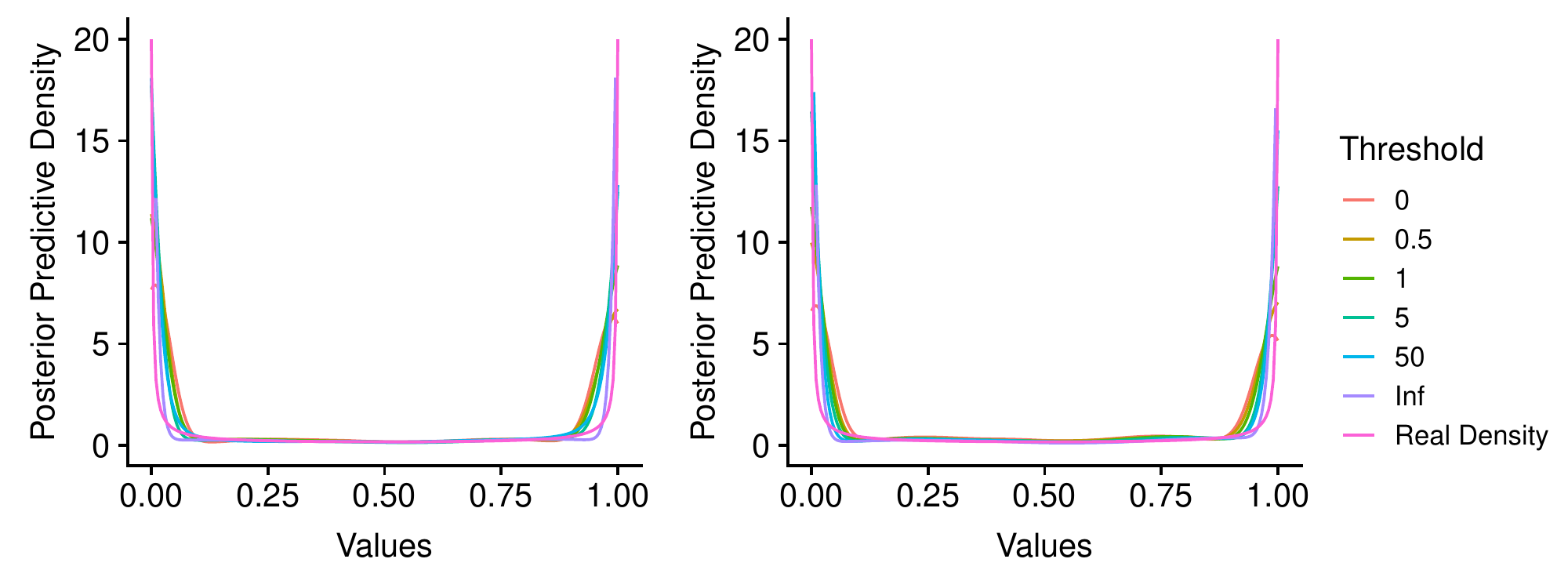}
\end{minipage}
\vspace{-.1in}
\begin{minipage}{0.25\textwidth}
    \caption{{Posterior } density of MoTG (left) and TMoG (right) 
    for data from a Beta$(0.1, 0.1)$.}
    \label{fig:post_pred_beta}
\end{minipage}
\begin{minipage}{0.7\textwidth}
    \centering
    \includegraphics[scale=.6]{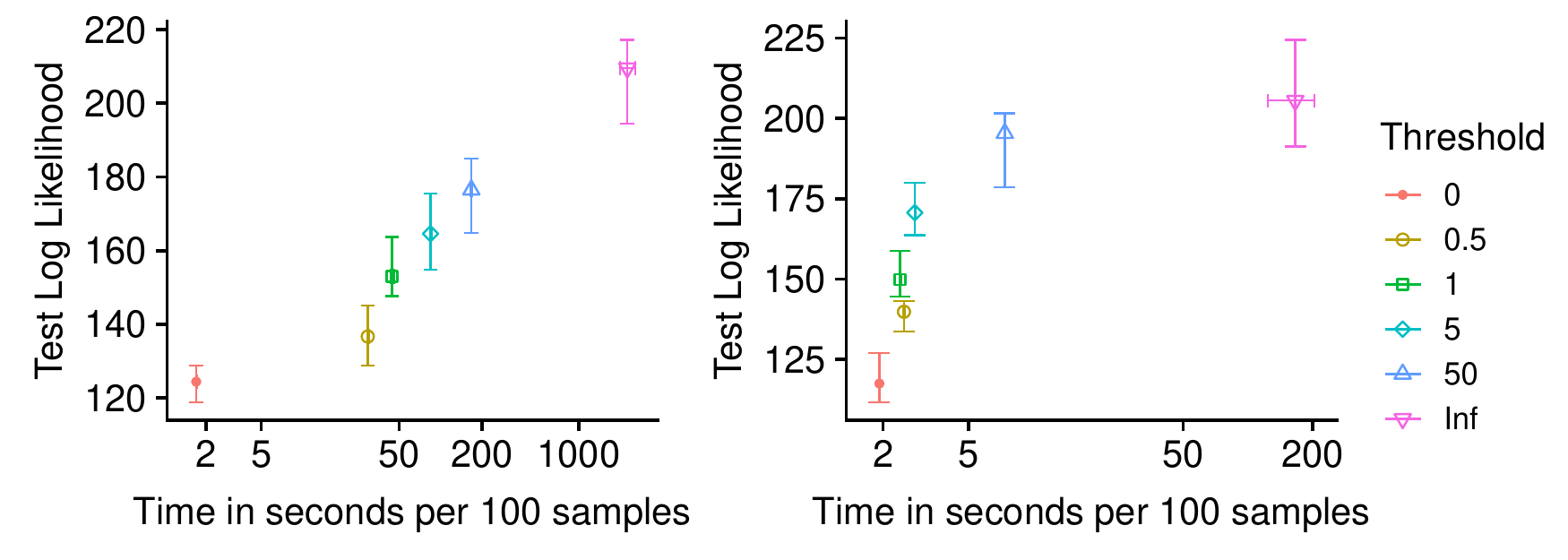}
\end{minipage}
\begin{minipage}{0.25\textwidth}
    \caption{Speed-accuracy plot for various thresholds for MoTG (left) and TMoG (right) for data from Beta(0.1, 0.1)}
    \label{fig:betaperf_plot_tmog_motg}
\end{minipage}
\end{figure}

The previous section suggested that the number of auxiliary rejected samples need not be much larger that the observed dataset; in fact we recommend setting $\rho=0.5$ (corresponding to a threshold of 1) for typical settings. Our next experiment, while still simple, considers a situation where larger thresholds are required. Now, we used TMoG and MoTG to model observations from a Beta$(0.1, 0.1)$ on the interval $[0,1]$. 
This parameter setting results in sharp modes at each end of the interval. 

Figure \ref{fig:post_pred_beta} plots estimated {} densities for both MoTG and TMoG. Like the previous experiment, we see that if the truncation is too strong, the model fails to adequately capture the peaks at the boundaries of the interval. Looking at the quantitative results in~\ref{fig:betaperf_plot_tmog_motg}, we see a clear monotonic improvement in test-likelihood for both MoTG and TMoG as more and more rejected samples are included. 
Now, capturing the sharp peaks at the edges of the boundary requires more rejected samples outside the interval: from this figure, it is clear that thresholding below 5 rejections per observation results in poor performance, with no thresholding performing best.
Of course, in practice, one does not know the true density.
In such situations, one can use the number of observations near the edges as a guideline for setting the threshold.
As before, this accuracy comes at a computational cost, with larger thresholds having longer run-times. 
Again, each TMoG setting is significantly more efficient than its MoTG counterpart.

\subsection{A 2-dimensional problem}
{In this section, we consider a significantly more complex 2-dimensional domain $\sub$, shown in figure~\ref{fig:2_contour_plot_bvgauss_poly}. 
We randomly draw 500 observations from 3 bivariate Gaussians, centered on the edges of an island. 
We split these into training and test sets of size 400 and 100, and then run 5000 MCMC interations with a burn-in of 2000. 
For both TMoG and MoTG, we place Normal-Inverse-Wishart priors on the components: 
 \begin{align}
   \mu | (\mu_0, \lambda, \Sigma) \sim N(\mu_0, \Sigma/\lambda), \quad \Sigma|(\Phi, \nu) \sim \text{Inv-Wish}(\Phi, \nu).
 \label{eq:prior_niw}
 \end{align}
 The parameters are: $\mu_0 = (0.5,0.5)$, $\lambda = 0.1$, 
 $\Phi = 0.001I_2$, and $\nu = 4$, where $I_2$ is the 2-dimensional identity matrix. 
 Figure \ref{fig:2_contour_plot_bvgauss_poly} shows the estimated density for TMoG for different $\rho$, and again we see a steep drop in density outside the constraint set, which is likely to lower probability density at the boundaries. 
 Figure~\ref{fig:perf_plot_bvgauss_poly} plots performance of TMoG and MoTG for different $\rho$: again, TMoG is much faster. 
Further, TMoG's predictive performance plateaus at thresholding equal to 1 ($\rho=0.5$), while MoTG requires higher thresholds. 
That a threshold of 1 does a good job can be guessed from the fact that unlike the beta distribution, observations are relatively well-dispersed near the boundaries.

\begin{figure}[H]
\centering

\includegraphics[width=0.3\textwidth]{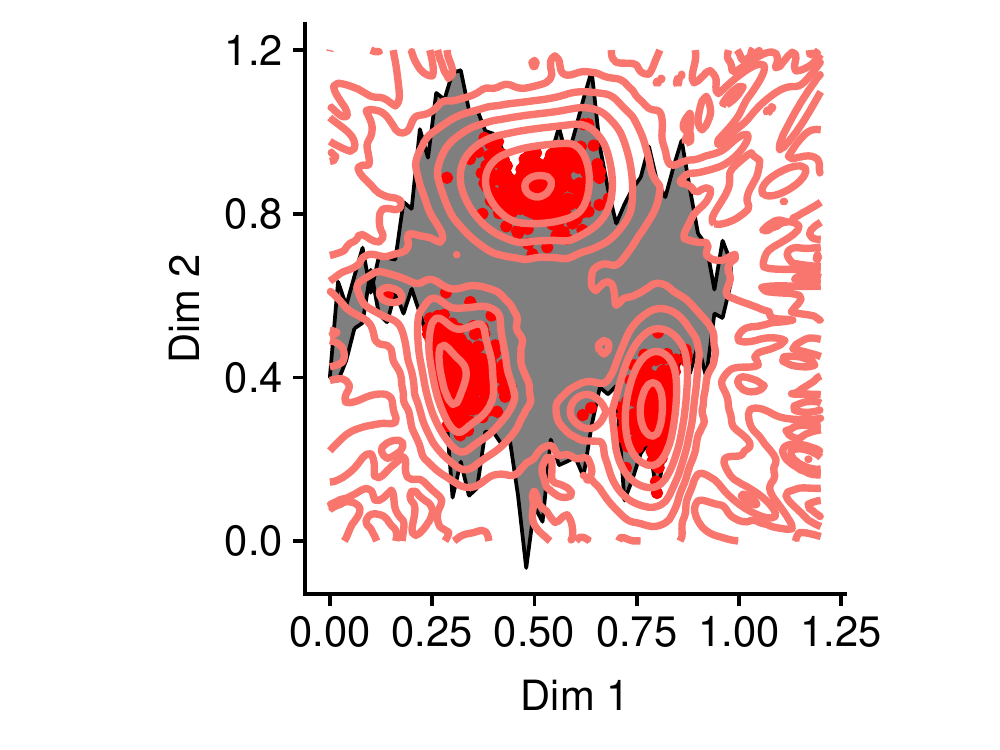}
\includegraphics[width=0.3\textwidth]{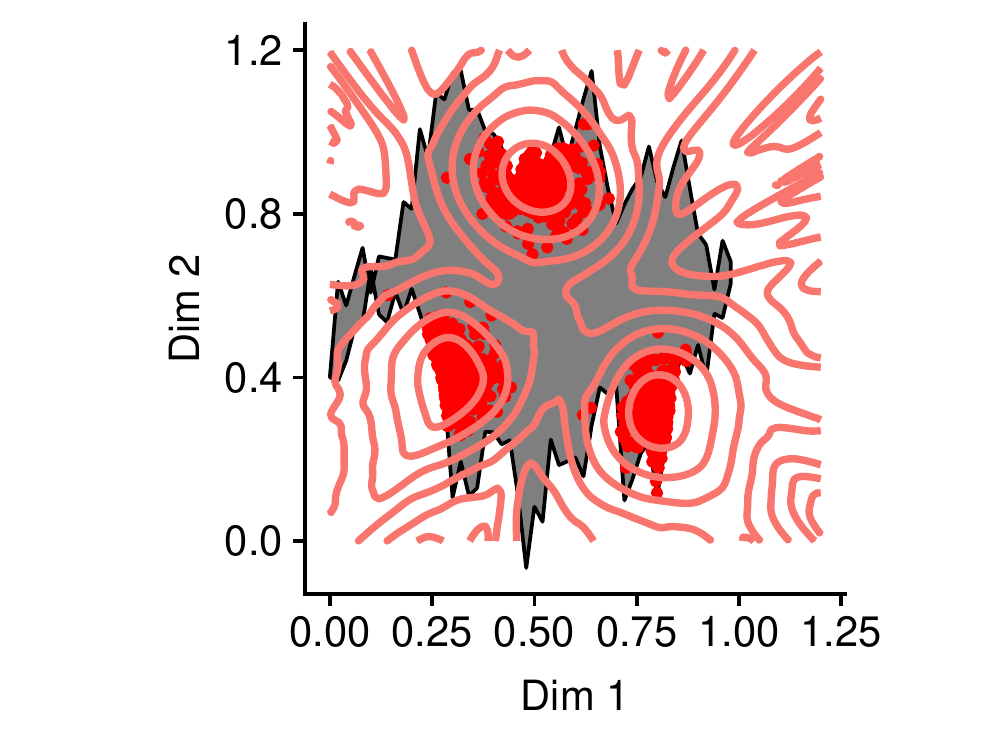}
\includegraphics[width=0.3\textwidth]{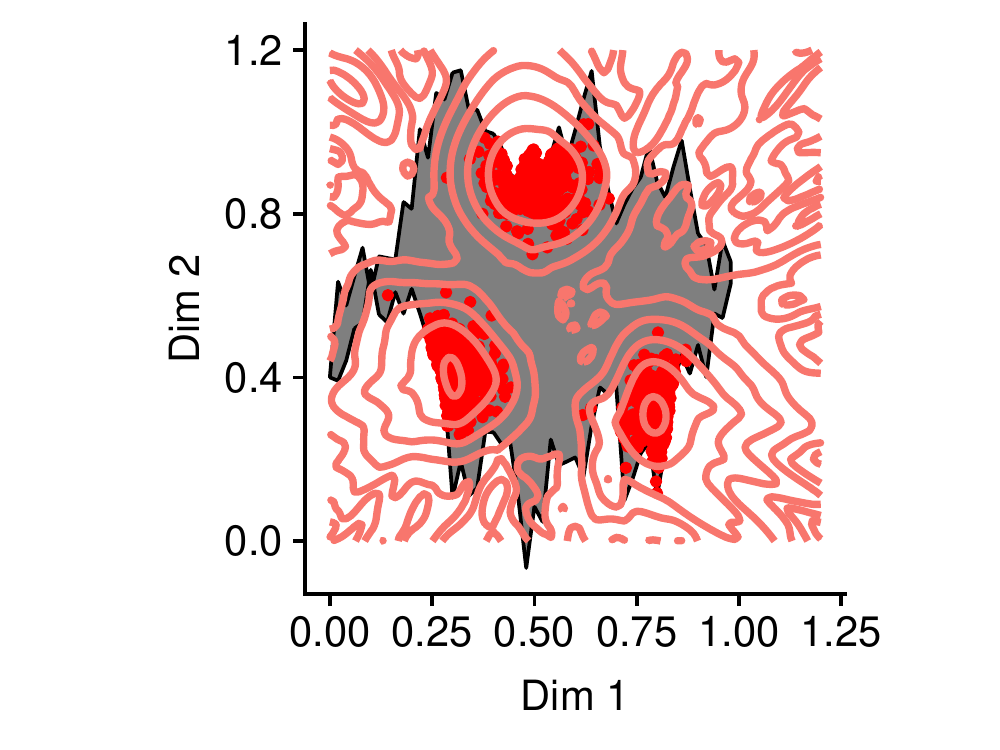}
\includegraphics[width=0.3\textwidth]{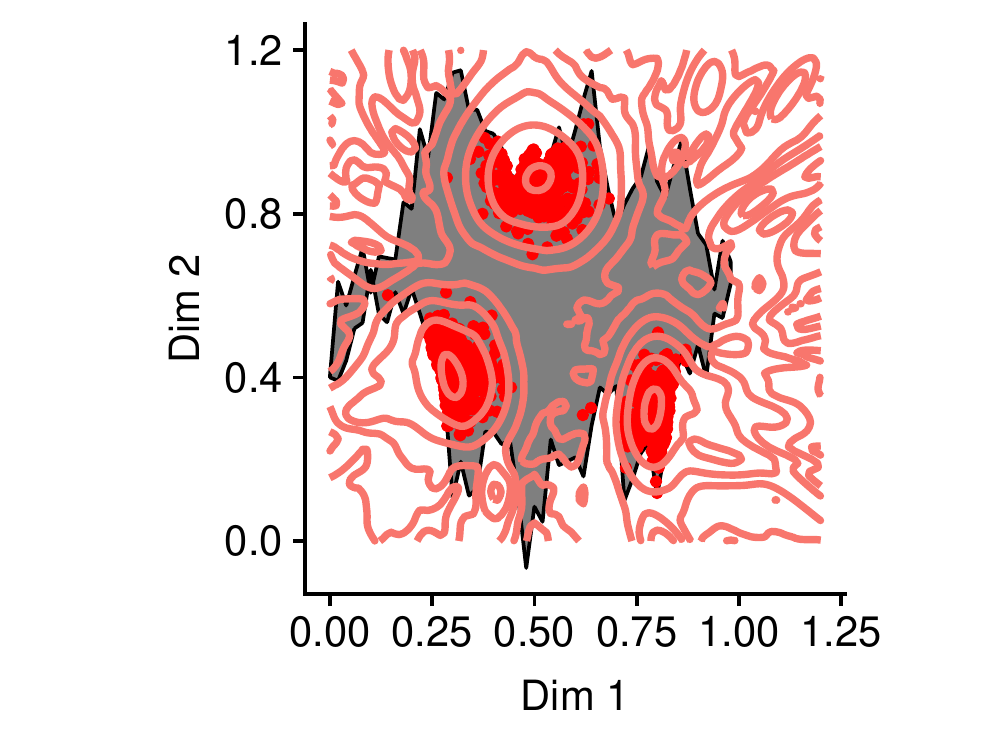}
\includegraphics[width=0.3\textwidth]{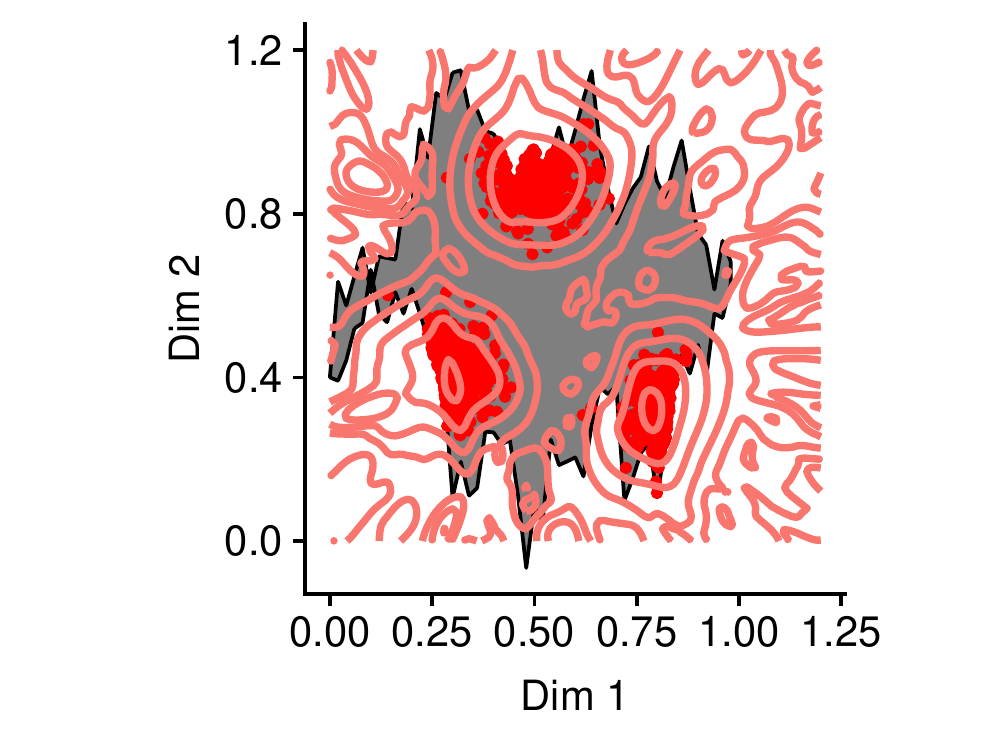}
\includegraphics[width=0.3\textwidth]{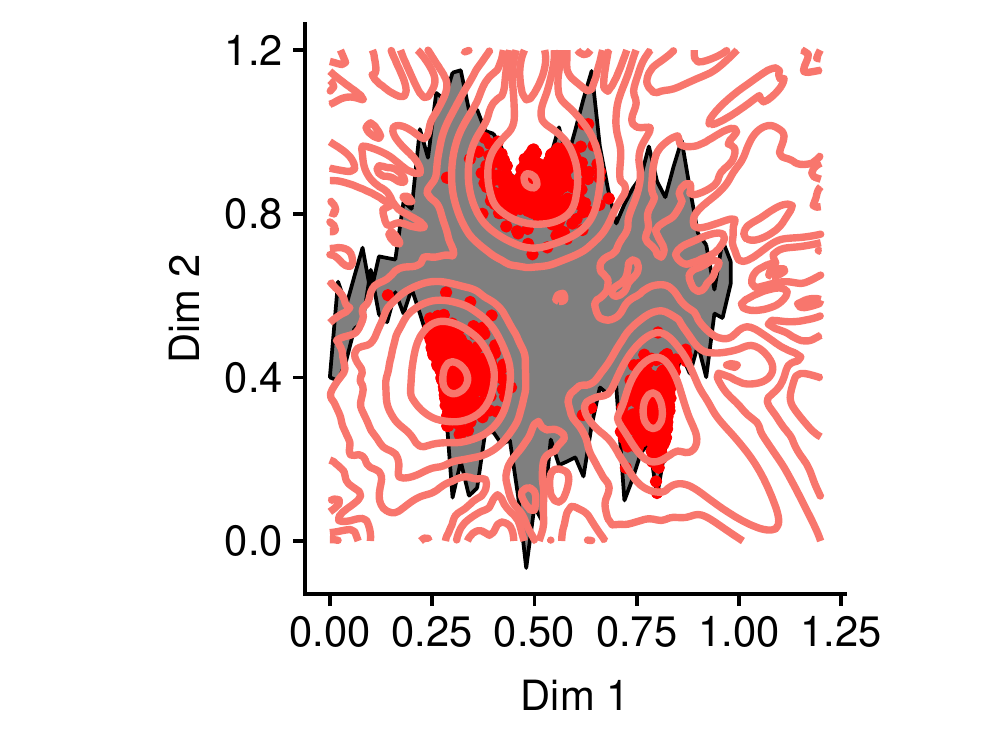}
\caption{Contour plot of log posterior mean density for TMoG for different $\rho$.
Thresholds are 0, 0.5, 1, 5, 50, and Infinity (left to 
right, top to bottom)}
\label{fig:2_contour_plot_bvgauss_poly}
\end{figure}

\begin{figure}[H]
\begin{minipage}{0.8\textwidth}
\centering
\includegraphics[width=0.98\textwidth]{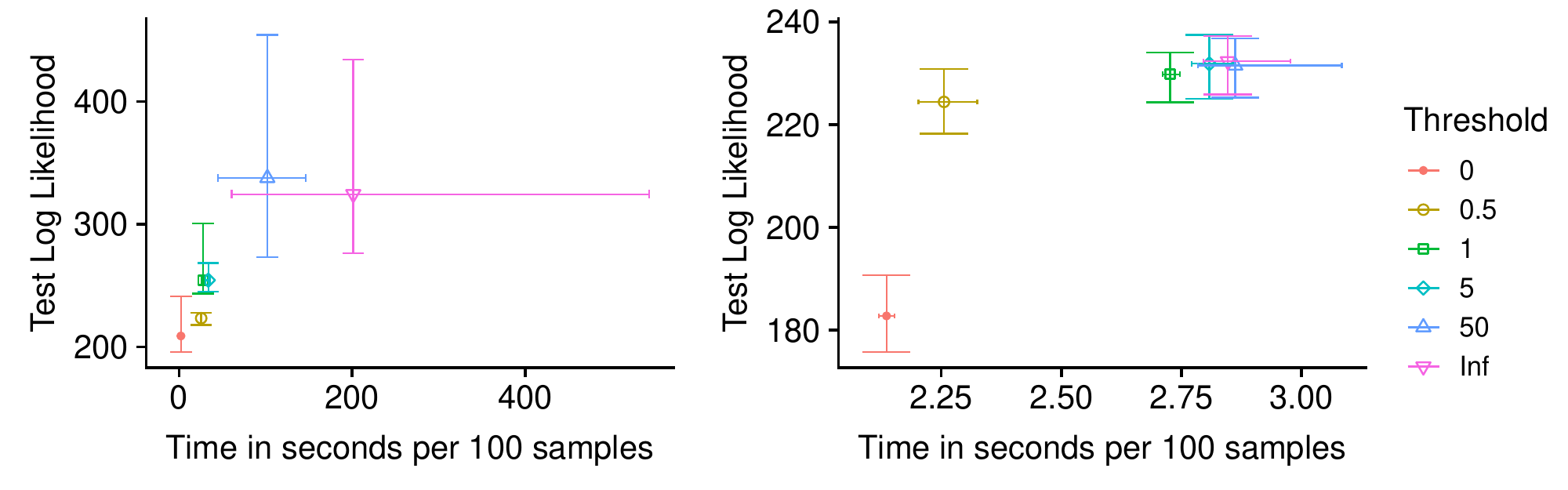}
\end{minipage}
\begin{minipage}{0.18\textwidth}
\caption{Speed-accuracy plots for MoTG (left) and TMoG (right).} 
\label{fig:perf_plot_bvgauss_poly}
\end{minipage}
\centering
\begin{minipage}{0.47\textwidth}
\includegraphics[width=0.8\textwidth]{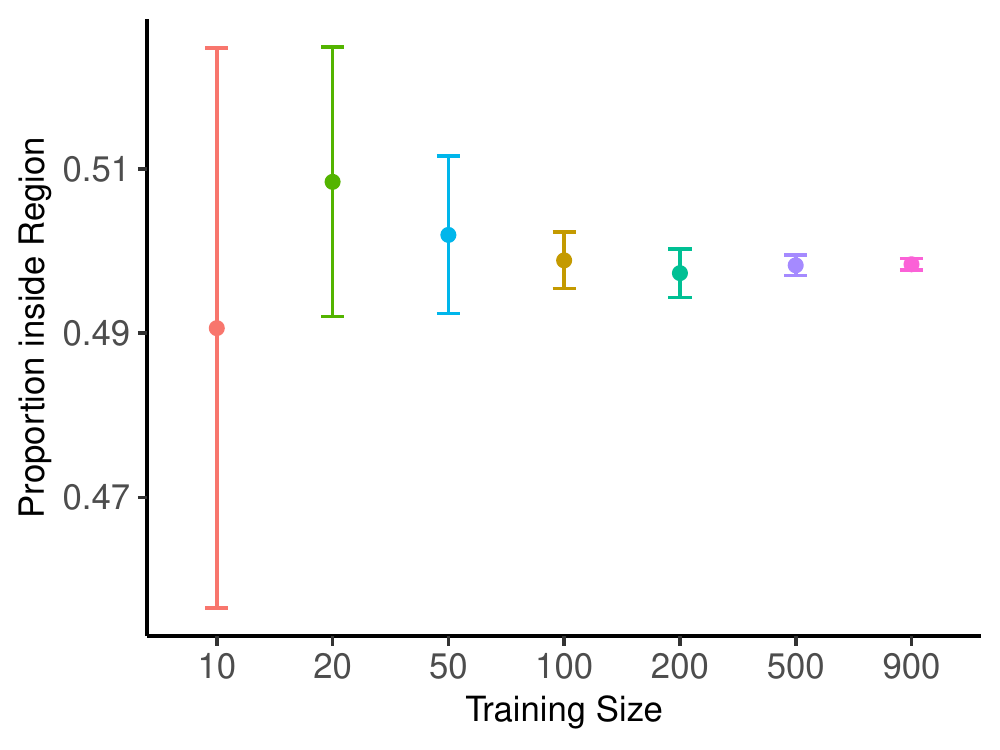}
\end{minipage}
\begin{minipage}{0.4\textwidth}
\caption{Estimates of $q(\sub)$ for $\rho = 0.5$ as training size increases. The MCMC sampler is run with a threshold of $1$, corresponding to $\rho=0.5$} 
\label{fig:tmog_expNum_prop_1}
\end{minipage}
\end{figure}

%

{Recall that our finite threshold sampler simulates from a model requiring 
  $q(\sub) \ge \rho$. 
  Our sampler is an approximate MCMC sampler though, since it does not enforce this constraint when updating the cluster parameters (step 2 in section~\ref{sec:thrdars}). 
We justified this by arguing that standard consistency results ensure that the approximation error is small with large sample sizes. 
Figure~\ref{fig:tmog_expNum_prop_1} verifies this empirically, plotting Monte Carlo estimates of $q(\sub)$ as 
the number of observations is increased. 
Here, the MCMC sampler was run with a threshold of $1$, and we see that the $q(\sub)$ we recover is close to $0.5$ for datasets larger than 100, justifying our approximation.
 
} 

\subsection{Flow Cytometry data}
\label{sec:app}

\begin{figure}[ht]
    \centering
        \includegraphics[scale=0.6]{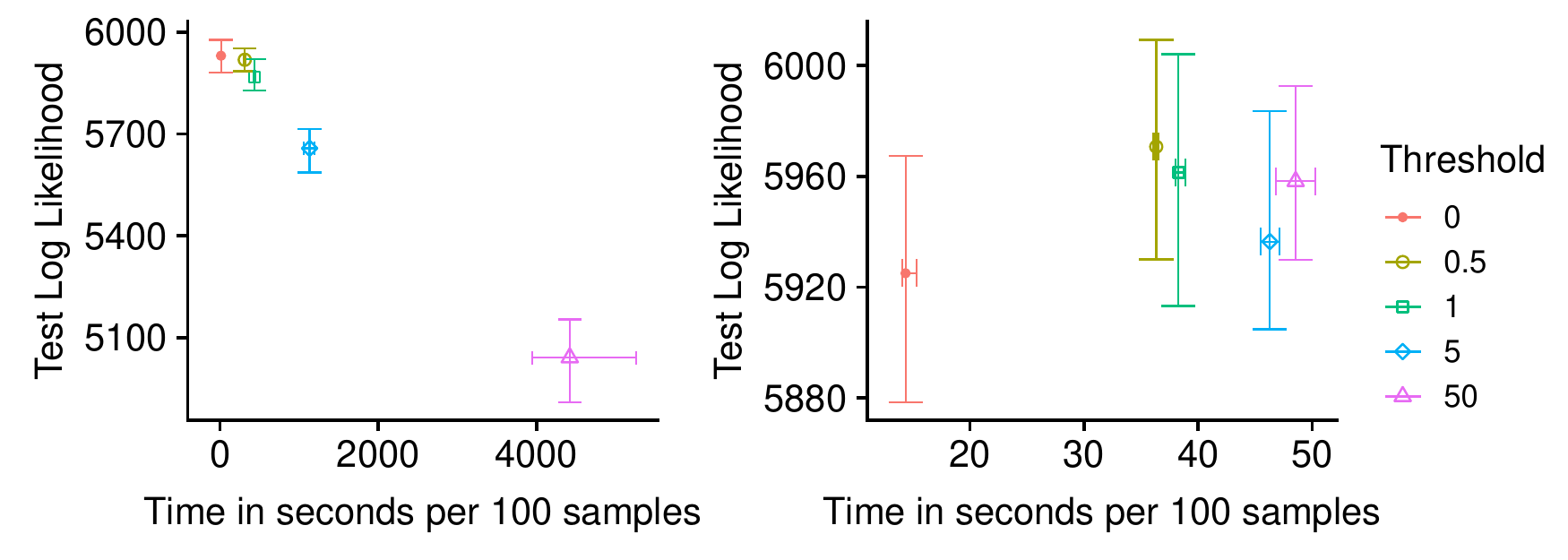}
    \caption{Speed vs accuracy for MoTG (left), and TMoG (right) 
    for the flow-cytometry data.}
    \label{fig:flowcytoperf}
\end{figure}
\begin{figure}[ht]
    \centering
        \includegraphics[scale=0.9]{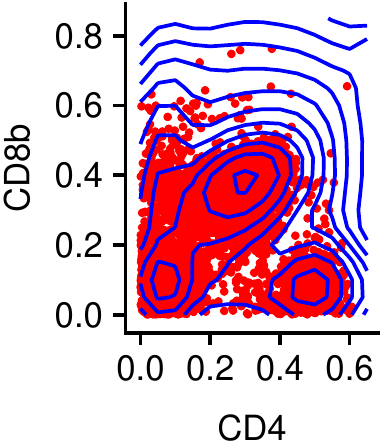}
        \includegraphics[scale=0.9]{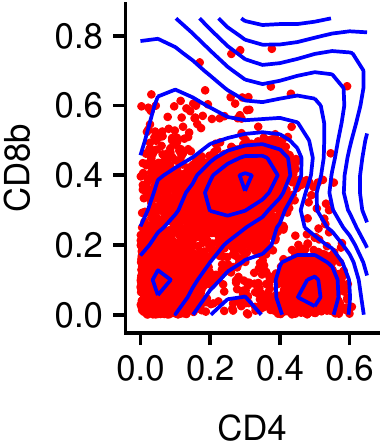}
        \includegraphics[scale=0.9]{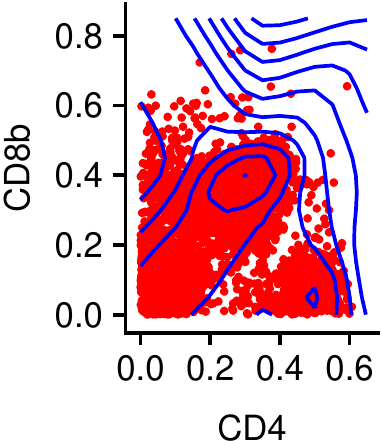}
        \includegraphics[scale=0.9]{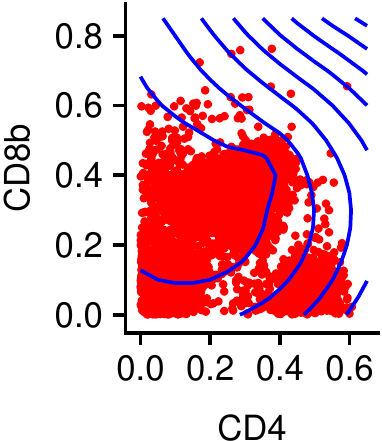}
        \includegraphics[scale=0.9]{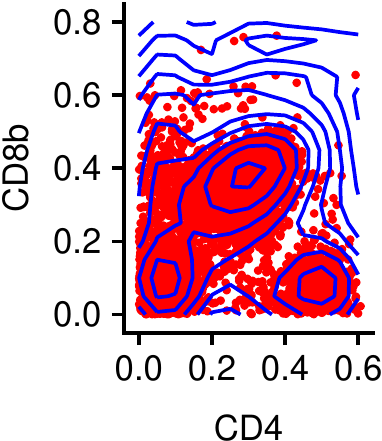}
        \includegraphics[scale=0.9]{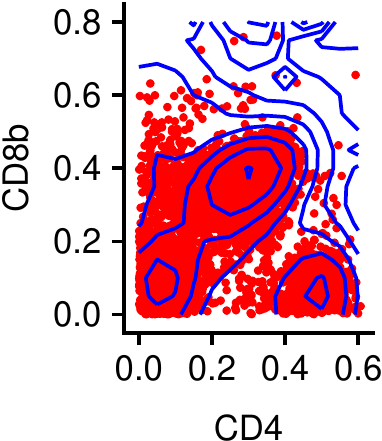}
        \includegraphics[scale=0.9]{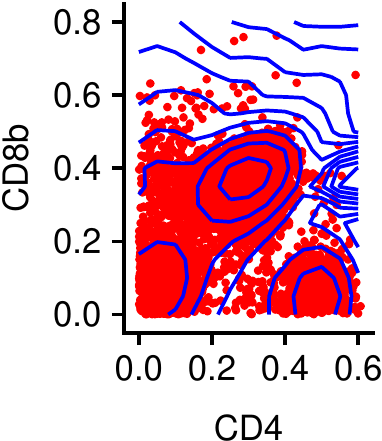}
        \includegraphics[scale=0.9]{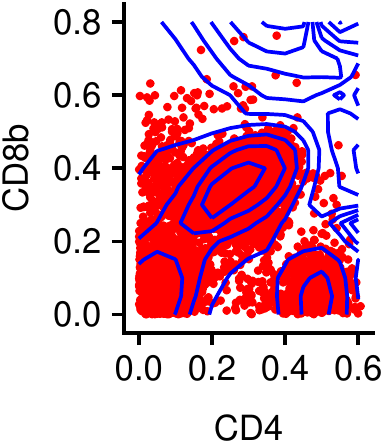}
   \caption{Contour plots of the posterior mean distribution for 
    MoTG (top) and TMoG (bottom) applied to flow-cytometry data. 
    Subfigures are thresholds of 0, 1, 5, and 50.} 
    \label{fig:cont_plot_flowcyto} 
\end{figure}
We next consider a dataset of acute graft-versus-host 
disease (GvHD) flow-cytometry measurements from patients with 
bone-marrow transplants \citep{brinkman2007high}. 
GvHD can arise  after receiving stem cells, when transplanted donor T-cells (``the graft'') attack the patient's 
healthy organs (``the host''). 
The dataset from~\citet{brinkman2007high} contains 6809 ``control'' and 
9083 ``positive'' observations from 31 patients. We focus on the control 
observations, which are from patients who did not develop either acute or 
chronic GvHD after the transplant. Each observation involves 
measurements of 4 activation markers: CD4, CD8b, CD3, and CD8, with each 
measurement varying between 0 and 1024. The data collection process is 
such that observations outside this range are discarded, resulting in a 
sharp drop in intensity outside this set (see 
Figure~\ref{fig:cont_plot_flowcyto} for projections of the raw data onto 
two-dimensional planes). We scale the data to lie in the 
four-dimensional unit hypercube, which forms our constraint set $\sub$. 

This dataset was briefly considered in~\citet{rao2016data}, where 
the authors demonstrated the feasibility of MCMC inference for TMoG. Here 
we analyze it more systematically, evaluating the performance of our 
two models, as well as different threshold settings. As before, in our 
mixture models we use the Normal-Inverse-Wishart as the prior, with 
parameters: 
$\mu_0 = 0.5$, $\lambda_0 = 0.01$, $\Phi = 0.001I_4$, and $\nu = 5$. 
$I_4$ is the four-dimensional identity matrix.

Figure \ref{fig:cont_plot_flowcyto} shows contour plots of the predictive 
density of the models for different threshold settings, while 
Figure~\ref{fig:flowcytoperf} shows how thresholding trades-off between 
predictive performance and compute time. 
For MoTG, we observe now that test-likelihood is no longer monotonic with the threshold setting, in fact, performance {\em worsens} as the threshold increases above 5.
This is largely because now we are working with 4-dimensional observations, and without controlling the prior over $q(\sub)$, it is easy to explore elements that produce very large numbers of rejected samples. 
This can severely affect MCMC mixing.
In fact, we do not include the case where the threshold is set to infinity, since this occasionally produced a very large number of rejections that brought our simulations to a near halt. 
Figure~\ref{fig:flowcytoperf} shows a steep drop in test log-likelihood for large thresholds, and in the contour plots of Figure~\ref{fig:cont_plot_flowcyto}, this manifests itself in the loss of multimodal structure. 
Figure \ref{fig:tp_flowcyto} shows MCMC traceplots of the number of rejected samples over MCMC iterations for MoTG, and we see that other than for the low thresholds, the MCMC chain mixes very poorly.
By thresholding the number of rejections, we are implicitly focusing on simpler model structure outside $\sub$, resulting in improved MCMC performance.

\begin{figure}[ht]
    \centering 
    \includegraphics[width=\textwidth]{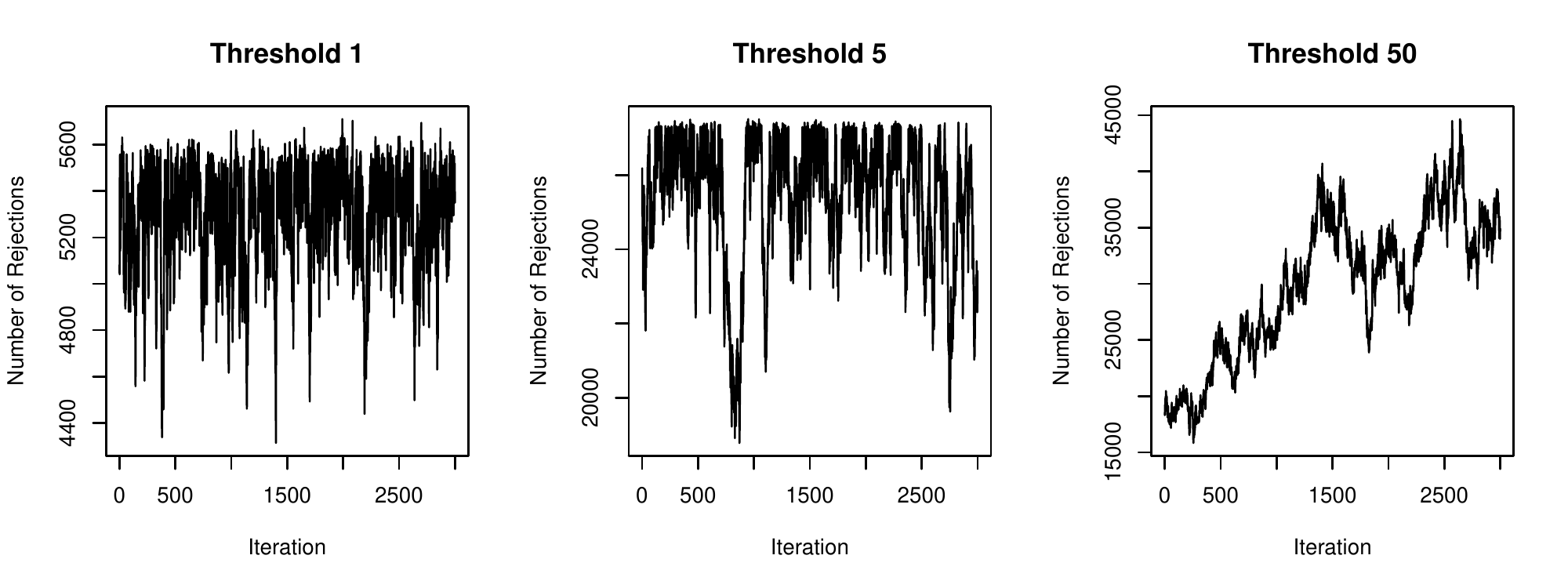}
    \caption{MCMC trace plot of number of rejections for MoTG for 
    flow-cytometry data}
    \label{fig:tp_flowcyto}
\end{figure}

The TMoG models run much faster, and produce much better fits than their MoTG counterparts; this is clear qualitatively from Figure \ref{fig:cont_plot_flowcyto} and quantitatively from Figure~\ref{fig:flowcytoperf}. 
We do not see any significant changes in predictive performance with increasing threshold (although there is improvement in median performance). 
More importantly we do not see performance decay as the threshold increases, though without any thresholding, we did observe occasional MCMC iterations with a large number of rejections. 
A quick analysis of the dataset reveals that less that 5\% of the observations lie within $0.01$ of an edge, suggesting (as the figure verifies) that a threshold of $1$ will be adequate.
For all threshold settings, TMoG demonstrated reasonable mixing (like the left-most plot in Figure~\ref{fig:tp_flowcyto}), with autocorrelations not extending beyond 10-15 iterations.

\subsection{Chicago crime data}
In our final experiment, we consider a dataset of criminal activity recorded in the city of Chicago. 
We gathered data from the city of Chicago data-portal\footnote{\url{https://data.cityofchicago.org/Public-Safety}}, 
and plot it in Figure~\ref{fig:examples}. 
Each recording in this dataset includes details such as case number, time, type of crime, as well as the longitude and latitude of the crime location. 
We restrict ourselves to modeling locations of homicide crimes occuring between the years 2012 to 2017, resulting in 3220 observations. 
Thus, each observation is a measurement in a two-dimensional space where the longitude and latitude are the x- and y-coordinates respectively. 
The range of longitude value is from $-87.8465$ to $-87.5316$, while latitude range from $41.6479$ to $42.0225$.  
We rescaled these values to between $-1$ and $1$. 
\begin{figure}[ht]
\centering
\includegraphics[width=0.3\textwidth]{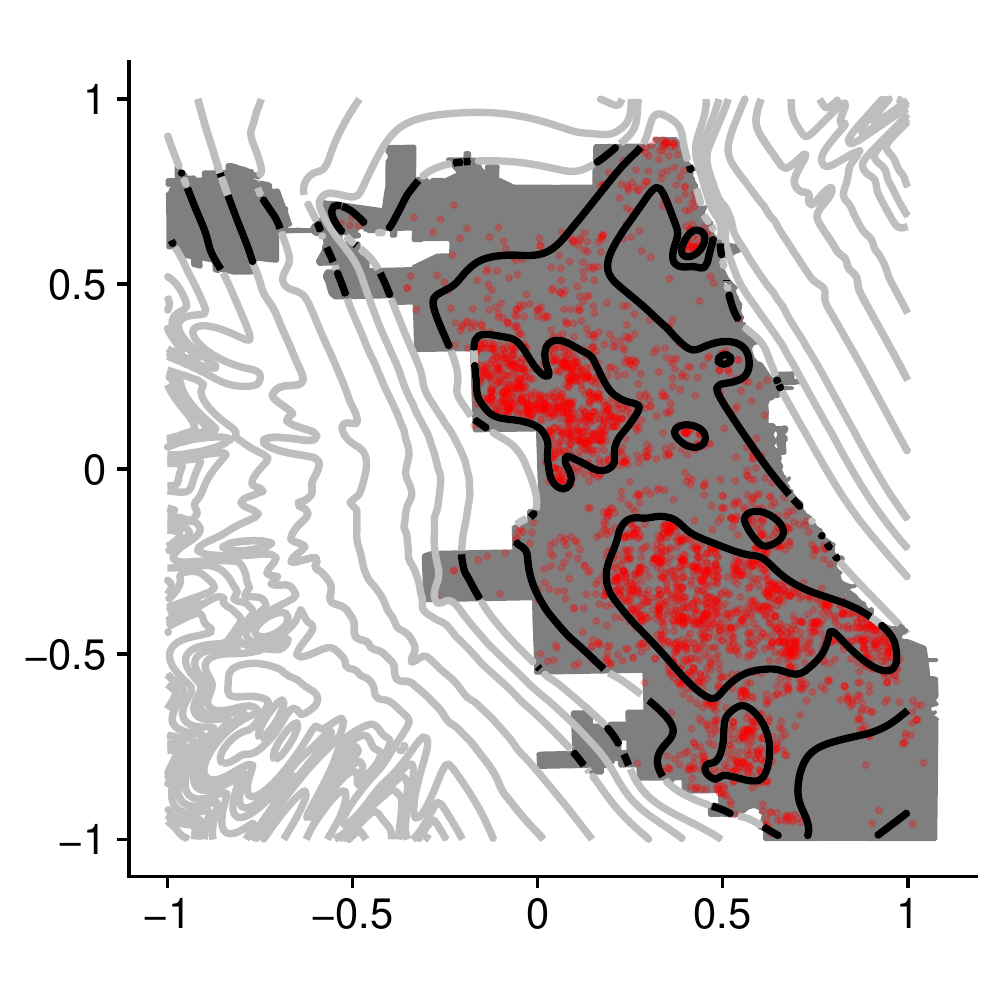}
\includegraphics[width=0.3\textwidth]{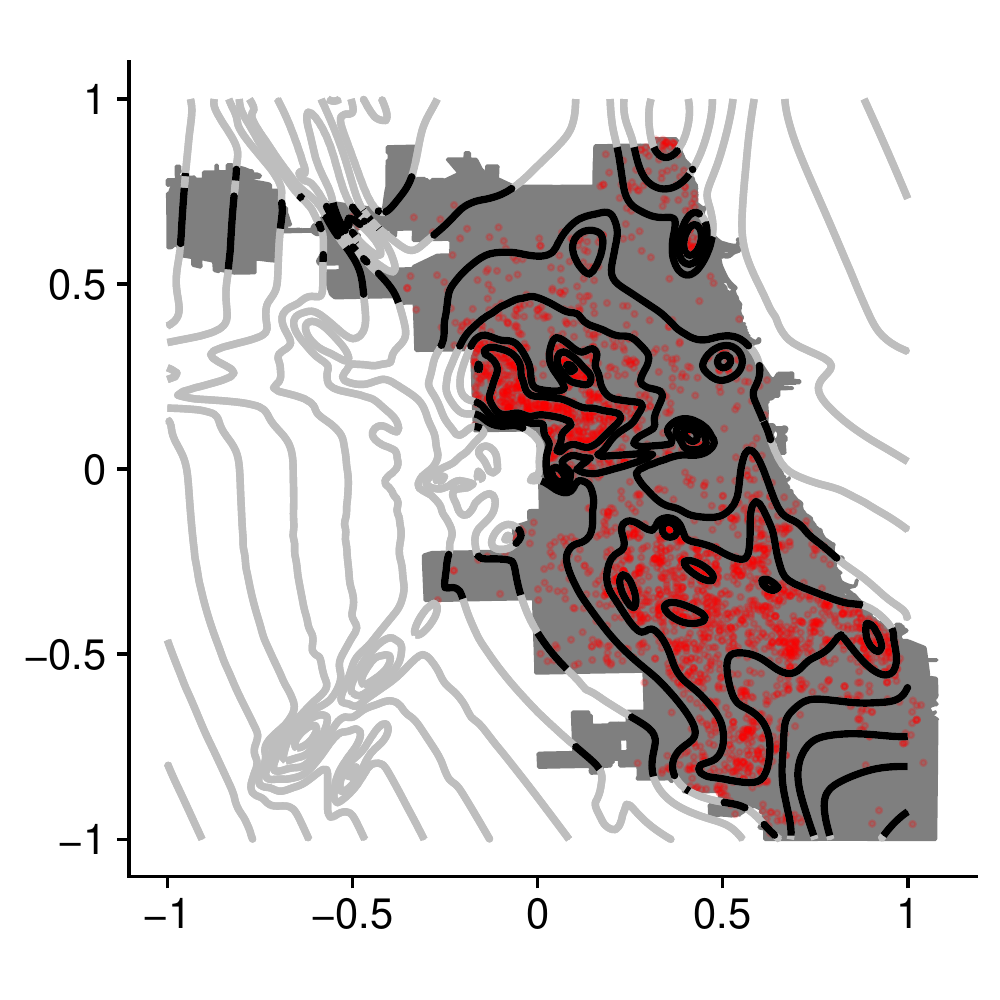}
\includegraphics[width=0.3\textwidth]{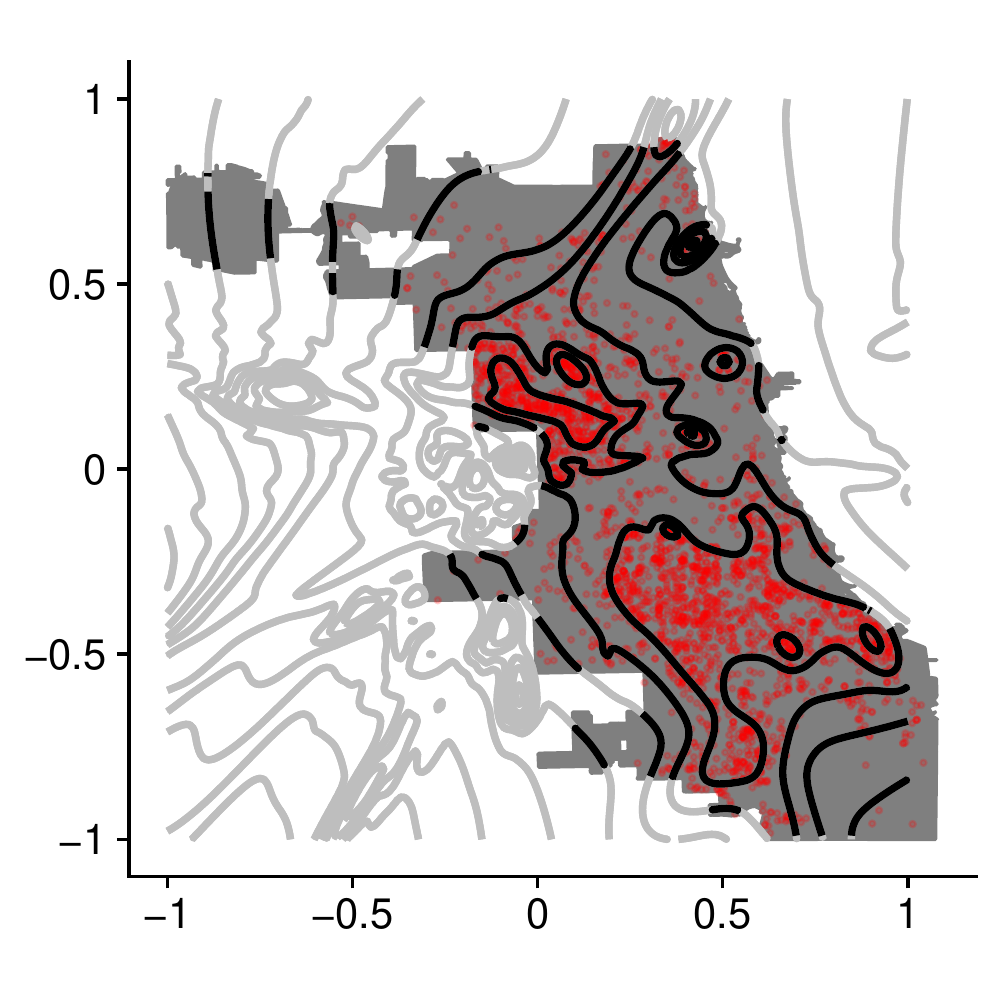}
\includegraphics[width=0.3\textwidth]{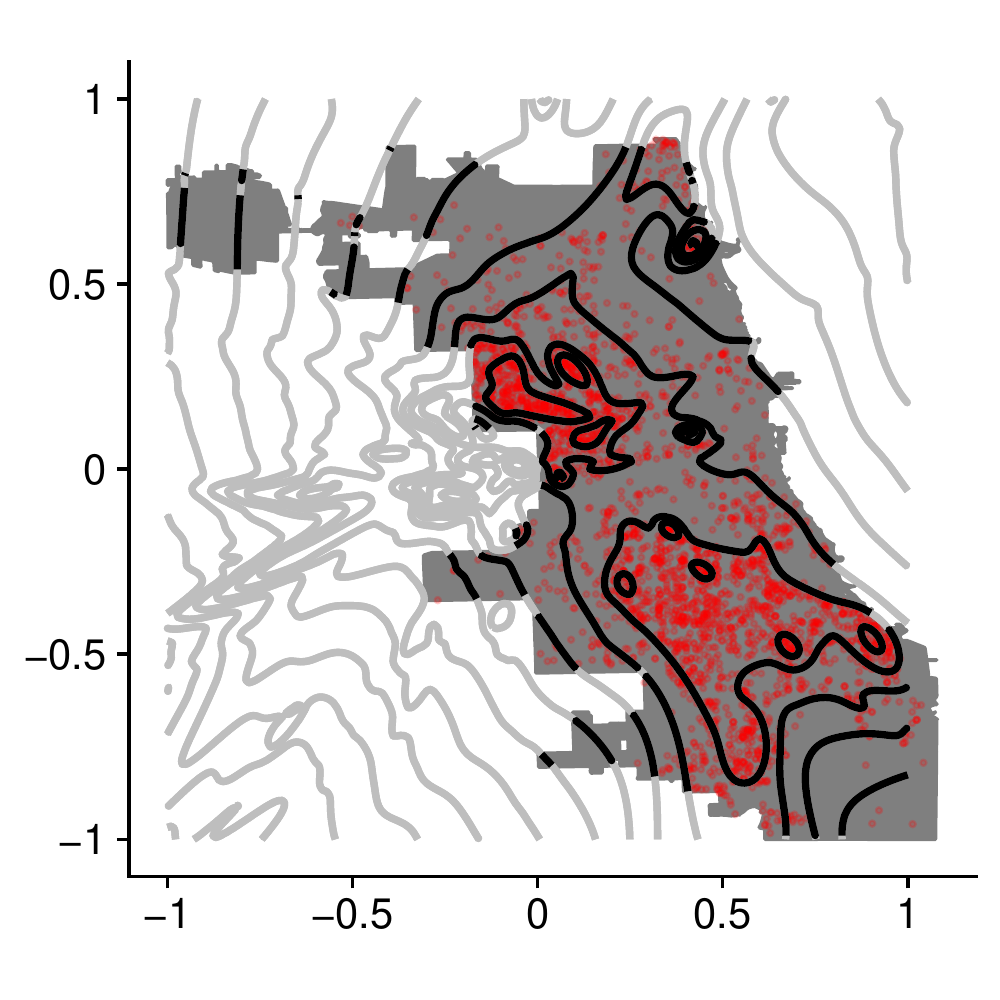}
\includegraphics[width=0.3\textwidth]{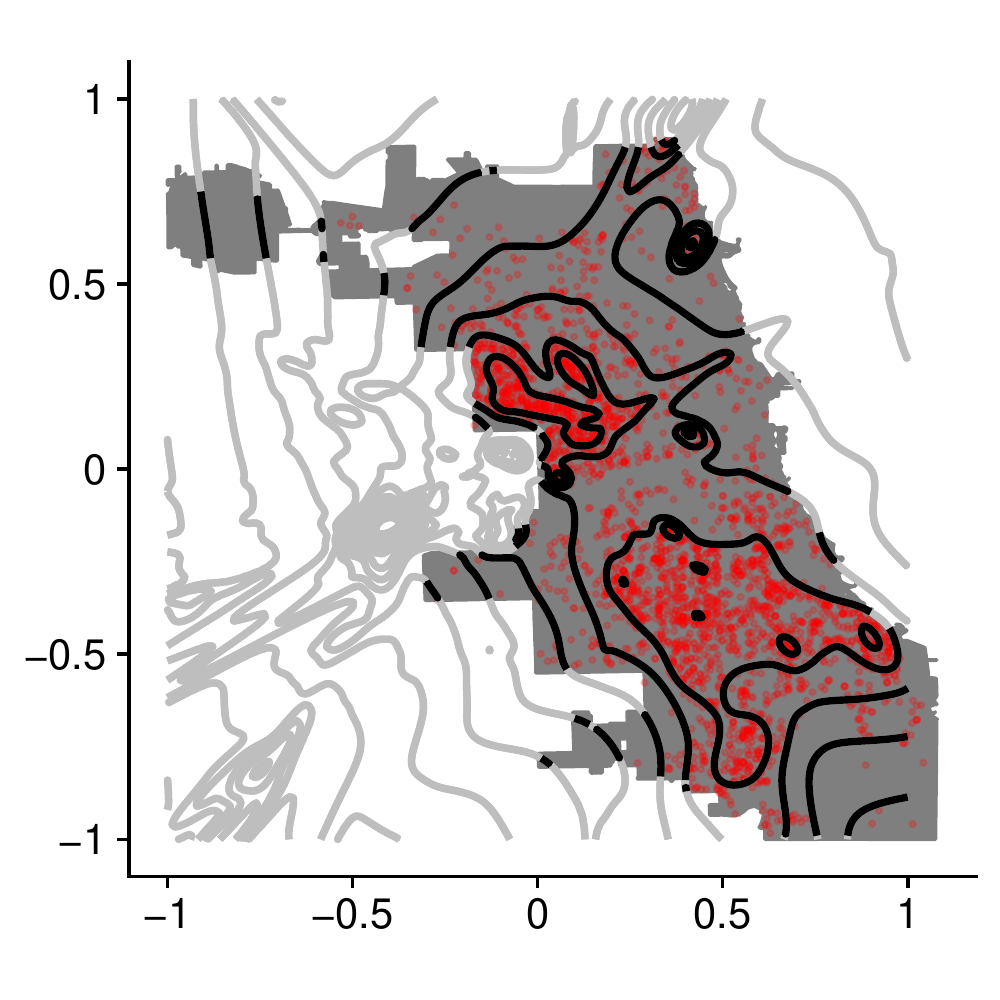}
\includegraphics[width=0.3\textwidth]{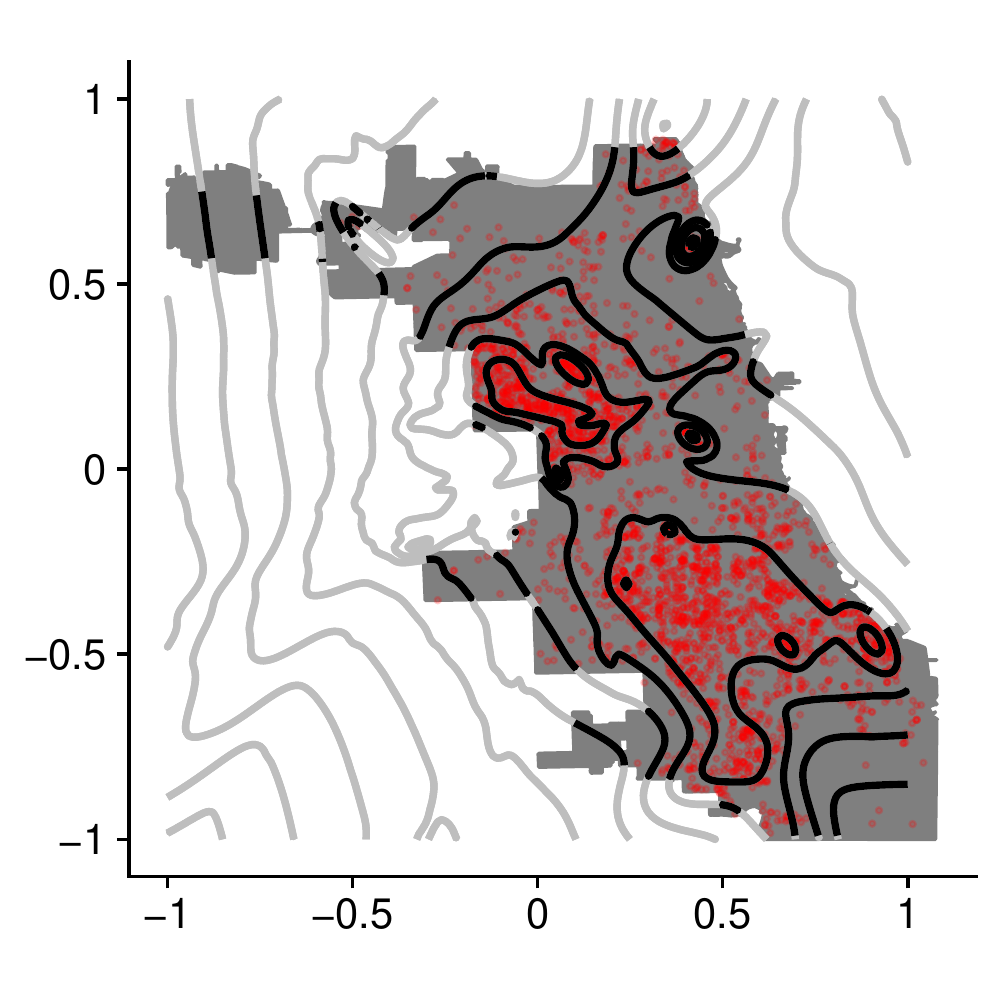}
\caption{Contour plots of the log posterior mean density for TMoG on the Chicago crime data with thresholds 0, 0.5, 1, 5, 50, and Infinity (left to right, and then top to bottom).}
\label{fig:tmogt_crime_contour_plot}
\end{figure}

Within the 2-dimensional Euclidean space, our constraint set $\sub$ is the interior of the city of Chicago. 
To characterize this complex, non-convex set, we gathered data for the boundaries of the 77 neighborhood limits of Chicago\footnote{\url{https://www.cityofchicago.org/city/en/depts/doit/provdrs/gis.html}}, 
and approximated each neighborhood with a polygon using the $\mathtt{R}$ spatial polygon package, $\mathtt{SP}$\footnote{\url{https://cran.r-project.org/web/packages/sp/}}.  
Combined together, these polygons formed the entire city limits. 
The package $\mathtt{SP}$ also allows to check whether a point lies inside a polygon. 
We used this function in our rejection sampling algorithm, to decide whether or not a proposal on $\mathbb{R}^2$ lies within Chicago. 
  
Below, we present results from modeling this data using TMoG. 
We do not include results from MoTG since, as indicated by our previous experiments, this takes much longer to run and produces results that 1) are much worse and 2) are sensitive to $\rho$. 
For our prior over cluster parameters, we used a two-dimensional Normal-Inverse-Wishart distribution with parameters $\mu_0 = (0,0)$, 
$\lambda_0 = 0.1$, $\Phi = 0.001 I_2$, and $\nu = 4$. 
Figure~\ref{fig:tmogt_crime_contour_plot} visualizes the posterior distribution through samples from the MCMC algorithm for different settings of the threshold parameter. 
Each subplot shows the log of the posterior mean density for TMoG given the crime data, with the grey contour lines showing the estimated proposal distribution $q$, and the black lines highlighting them within the Chicago limits. 
The latter gives (up to a constant) the density of interest, whose properties can easily be estimated from the MCMC samples. 

For all threshold settings, we observe two modes, one to the south of Chicago, and one to the west. 
When the threshold is set to $0$, the range of density values near the boundary is smaller, being flattened to account for the absence of observations just outside the border. 
This misses many details near the boundary, for instance, there is a sharp cluster of observations right on the north-east boundary of Chicago which is lost for the threshold settings of $0$. 
For larger threshold settings, the rejected proposals produce a significant cluster most of whose mass lies outside the city limits, but which overlaps with the city to allow a bump in probability at the corner. 
Edge smoothing-effects due to the constraint also result in coarser estimates at the mode to the west of the city. 

An interesting phenomenon is the mild structure in the density contours 
away from the city boundaries. These are transients, resulting from the 
data-augmentation interacting with the thresholding. These do not 
affect inferences over the subset of interest, and are not relevant to 
the main estimation problem. Nevertheless, these represent an inefficiency 
in the thresholding procedure, and a waste of computational resources. A 
future research direction is to favor thresholding away from the subset $\sub$ of interest.

\begin{figure}
\begin{minipage}{0.74\textwidth}
    \centering 
    \includegraphics[width=0.97\textwidth]{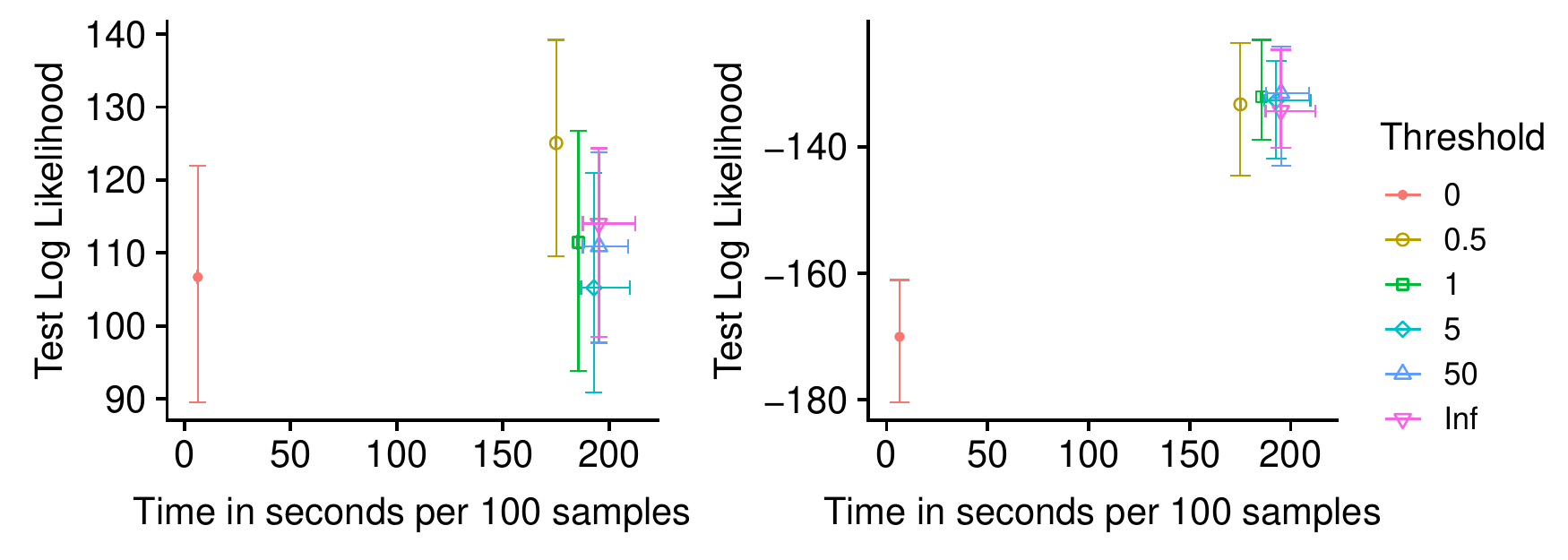}
  \end{minipage}
\begin{minipage}{0.24\textwidth}
    \caption{Speed-accuracy plot of TMoG on Chicago test data. The left figure gives the performance on test data with random observations, the right figure gives the performance on test data with observations near the boundary.}
    \label{fig:perf_plot_crime_tmogt}
  \end{minipage}
\end{figure}

Figure~\ref{fig:perf_plot_crime_tmogt} quantifies the effect of 
threshold settings, plotting predictive performance of TMoG on held-out test data versus 
threshold. The left plot gives results for test data 
randomly sampled from the original dataset, and we see a slight, but not significant 
performance hit with no augmentation. 
The right subplot presents results when the test dataset is drawn from observations 
in Chicago neighborhoods touching the boundary. Now we see a 
significant loss of predictive performance without data-augmentation, 
providing quantitative justification of the importance 
of our data-augmentation scheme to accurately model probability structure 
near the edges of the constraint set. 

As far as run-time is concerned, we see that a threshold of $0$ is 
much faster that other settings, and that the average run-time 
does not increase significantly for larger thresholds. In fact, 
the relative inefficiency of these settings has little to do with our 
data-augmentation scheme. For instance, our recommended setting of TMoG, 
with threshold set to $1$ produces on average around 2346 rejected proposals, which is less than a 50\% increase in the original dataset.
Instead, the increased run-time reflects our relatively crude 
approach to deciding if a proposal lies in the Chicago city limit. 
Our implementation, using the \texttt{R} package \texttt{SP}, 
needs to check that a proposal does not lie in any of the $72$ Chicago neighborhoods before we 
can reject it. This accounts for the bulk of the run-time for non-zero 
thresholds (the vanilla mixture model without data-augmentation is 
unaware of the borders of the city and therefore does not include such 
checks). With a more careful implementation of the Chicago border, we can reduce this 
overhead. 

\begin{figure}
    \centering
\begin{minipage}{0.45\textwidth}
    \includegraphics[width = 0.95\textwidth]{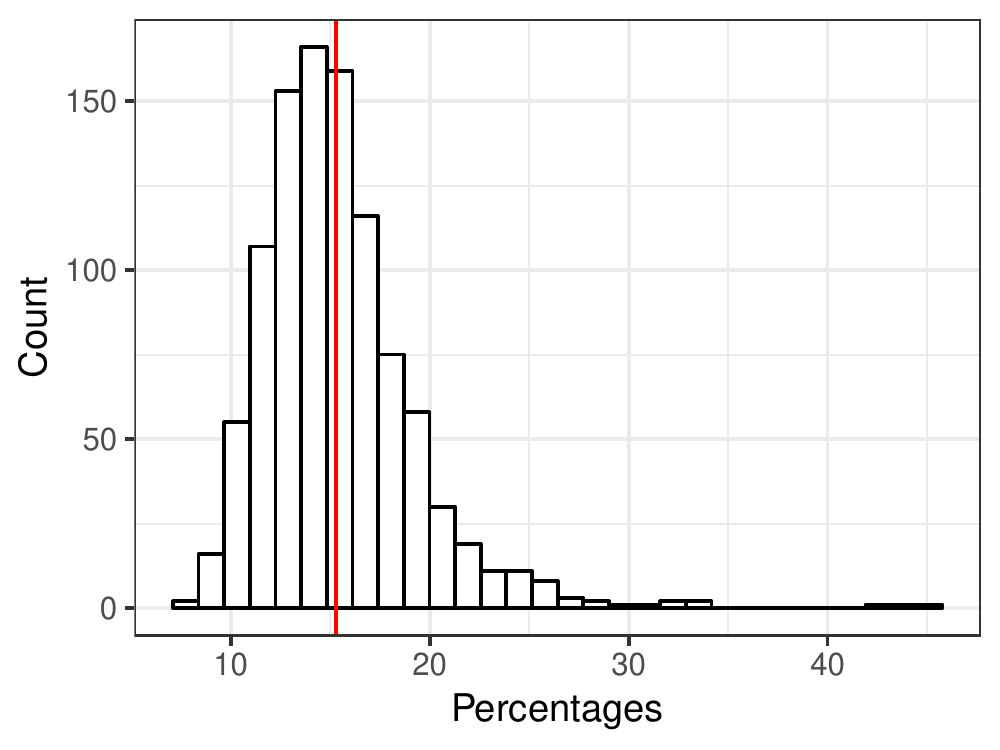}
  \end{minipage}
\begin{minipage}{0.34\textwidth}
  \caption{Distribution of percentages near the boundary from posterior predictive checks on 1000 MCMC samples for threshold 1. The vertical line at 15.28 gives the percentage of the real data with an estimated p-value of 0.456. } 
    \label{fig:postpredcheck}
  \end{minipage}
\end{figure}

To further validate our model, we used the MCMC posterior samples to run a posterior predictive check.
For each MCMC sample, we generated the same number of data points (3220 observations) inside Chicago. We then scaled the data by a factor of 1.2 and computed the percentage that lie outside of Chicago. We compare the distribution of this quantity to the realized value in the observed dataset. 
Figure \ref{fig:postpredcheck} plots both these quantities, with the observed value lying right in the middle of the predictive distribution. The resulting p-value (corresponding to the proportion of simulated values to the right of the observed quantity) was 0.46, suggesting a good fit.




\section{Discussion}
\label{sec:conc}
We proposed two approaches to modeling data lying in a contrained space: the truncated
mixtures of Gaussians (TMoG) and mixtures of truncated Gaussians (MoTG). Our methodology is most useful for 
nontrivial constraint sets, such as the boundaries of a city, though it is also useful in simpler settings like 
the simplex, or the unit disc, when existing mixture models are not flexible enough 
to represent rich correlation structure or to capture sharp changes in probability across boundaries. 
Implementation-wise, our code only requires an indicator function for the subset of interest, as well as a parameter $\rho$ related to how much probability the practitioner expects to see near the boundaries.
Our resulting MCMC algorithm is a very simple wrapper around standard sampling techniques. 

Future studies can extend our work to time-series models with restricted domains, and to a more 
complex constraint sets like manifolds. 
Our thresholding scheme serves to regularize the prior over the proposal distribution, limiting how much mass it assigns outside the constraint set. 
It is {interesting} to look at more refined approaches to this, such as increasing the likelihood of truncation with distance from the constraint set. 
Finally, it is of interest to better theoretically understand how bounding the probability of the proposal distribution $q$ on the subset $\sub$ can improve MCMC mixing. 

\bibliographystyle{apalike}
\bibliography{mixmod}

\end{document}